# The MACHO Project First Year LMC Results: The Microlensing Rate and the Nature of the Galactic Dark Halo

C. Alcock[1,2], R.A. Allsman[3], T.S. Axelrod[1,4], D.P. Bennett[1,2],
K.H. Cook[1,2], K.C. Freeman[4], K. Griest[2,5], J.A. Guern[2,5],
M.J. Lehner[2,5], S.L. Marshall[2,6], H.-S. Park[1], S. Perlmutter[2], B.A. Peterson[4],
M.R. Pratt[2,6], P.J. Quinn[4], A.W. Rodgers[4], C.W. Stubbs[2,6,7], W. Sutherland[8]
(The MACHO Collaboration)

1 : Lawrence Livermore National Laboratory, Livermore, CA 94550
Email: `alcock, bennett, kcook, hyesook@igpp.llnl.gov`

2 : Center for Particle Astrophysics, University of California, Berkeley, CA 94720
Email: `saul@lbl.gov`

3 : Supercomputing Facility, Australian National University,
Canberra, ACT 0200, Australia
Email: `robyn@macho.anu.edu.au`

4 : Mt. Stromlo and Siding Spring Observatories,
Australian National University, Weston, ACT 2611, Australia
Email: `tsa@macho.anu.edu.au` ; `kcf,peterson,pjq,alex@merlin.anu.edu.au`

5 : Department of Physics, University of California, San Diego, CA 92039
Email: `kgriest,jguern,mlehner@ucsd.edu`

6 : Department of Physics, University of California, Santa Barbara, CA 93106
Email: `mrp,stuart@lensing.physics.ucsb.edu`

7 : Departments of Astronomy and Physics, University of Washington, Seattle, WA 98195
Email: `stubbs@astro.washington.edu`

8 : Department of Physics, University of Oxford, Oxford OX1 3RH, U.K.
Email: `w.sutherland@physics.ox.ac.uk`

# Abstract

Since July 1992, the MACHO project has been carrying out long-term photometric monitoring of over 20 million stars in the Magellanic Clouds and Galactic Bulge. Our aim is to search for the very rare gravitational microlensing events predicted if the dark halo of our Galaxy is comprised of massive compact halo objects (hereafter Machos). We have now analysed most of the first year's LMC data, comprising 9.5 million light curves of stars with an average of 235 observations each. Automated selection procedures applied to this sample show 3 events consistent with microlensing, of which one is very striking (Alcock *et al.* 1993) and two are of modest amplitude. We have evaluated our experimental detection efficiency using a range of detailed Monte-Carlo simulations, including addition of artificial stars to real data frames. Using a 'standard' halo density profile we find that a halo comprised entirely of Machos in the mass range $3 \times 10^{-4}$ to $0.06 \, M_\odot$ would predict $> 15$ detected events in this dataset, and objects around $3 \times 10^{-3} \, M_\odot$ would predict 25 events; thus a standard spherical halo cannot be dominated by objects in this mass range. Assuming all three events are microlensing of halo objects and fitting a naive spherical halo model to our data yields a Macho halo fraction $f = 0.19^{+0.16}_{-0.10}$, a total mass in Machos (inside 50 kpc) of $7.6^{+6}_{-4} \times 10^{10} \, M_\odot$, and a microlensing optical depth $8.8^{+7}_{-5} \times 10^{-8}$ (68% CL). We have explored a wide range of halo models and find that, while our constraints on the Macho fraction are quite model-dependent, constraints on the total mass in Machos within 50 kpc are quite secure. Future observations from this and other similar projects and accurate measurements of the Galactic mass out to large radii should combine to give much improved constraints on the Macho fraction of the halo.

# 1. Introduction

As is well known, there is strong evidence for the existence of large amounts of dark matter surrounding the Milky Way and other spiral galaxies (Ashman 1992, Freeman 1995). This evidence is independent of theoretical arguments in favour of an $\Omega = 1$ universe. This dark matter cannot be in the form of low-mass stars, dust or gas, which would readily be detected (e.g. Bahcall *et al.* 1994). While a very wide range of candidates have been proposed, most of these candidates fall naturally into two classes: the 'particle physics' candidates such as axions, massive neutrinos or supersymmetric particles (Primack, Seckel & Sadoulet 1988), and the 'astrophysics' candidates such as sub-stellar 'Jupiters' and brown dwarfs below the hydrogen-fusion threshold $\sim 0.1 \, M_\odot$, or remnants of an early generation of massive stars, such as white dwarfs, neutron stars or black holes (Carr 1994). These astrophysical candidates are collectively known as massive compact halo objects or Machos.

While theoretical models of galaxy formation and microwave background anisotropies appear consistent with some models dominated by non-baryonic dark matter, it is clearly essential to detect the dark matter observationally to obtain a conclusive result.

If Machos exist, they would emit very little electromagnetic radiation; thus current optical and infrared searches are not sensitive enough to give useful constraints (Kerins & Carr 1994). However, Paczynski (1986) proposed that Machos could be detected by their gravitational 'microlensing' effect on background stars * . The principle is simple: if a Macho passes very close to the line of sight to a background star, the gravitational field of the Macho deflects the starlight and produces multiple images of the star. In the case of perfect alignment, the star will appear as an 'Einstein ring', with a radius in the lens plane of

$$r_E = \sqrt{\frac{4GmLx(1-x)}{c^2}} \tag{1}$$
$$= 2.85 \, \text{AU} \sqrt{\left(\frac{m}{M_\odot}\right)\left(\frac{Lx(1-x)}{1 \, \text{kpc}}\right)},$$

where $M$ is the lens mass, $L$ is the observer-star distance, and $x$ is the ratio of the observer-lens and observer-star distances. In a realistic case of imperfect alignment, the star will appear as two small arcs. For the scales of interest here, the image splitting is $\lesssim 0.001$ arcsec, and is far too small to be resolved; however, the multiple imaging results in an apparent amplification of the source (e.g. Refsdal 1964) by a factor

$$A = \frac{u^2 + 2}{u\sqrt{u^2 + 4}}, \tag{2}$$

where $u = b/r_E$ and $b$ is the separation of the lens from the observer-star line. Since objects in the Galaxy are in relative motion, this amplification will be transient, with a duration $\hat{t} \equiv 2\,r_E/v_\perp \sim 130\sqrt{m/M_\odot}$ days, where $v_\perp$ is the transverse velocity of the lens relative to the (moving) line of sight (Paczynski 1986, Griest 1991).

The "optical depth" ($\tau$) to microlensing is defined as the probability that any given star is microlensed with impact parameter $u < 1$ (i.e. $A > 1.34$) at any given time. Since $r_E \propto \sqrt{M}$, while for a given mass density, the number density is proportional to $M^{-1}$, the optical depth is independent of the mass function of Machos, and is given by

$$\tau = \frac{4\pi G}{c^2} \int_0^L \rho(l) \frac{l(L-l)}{L} dl, \tag{3}$$

where $\rho$ is the mass density of lensing objects and $l = Lx$ is the observer-lens distance. Using the virial theorem, it is readily shown that $\tau \sim (v/c)^2$ for a self-gravitating system with orbital velocity $v$, up to geometrical factors of order unity. For a "standard" spherical dark halo with density

$$\rho_H(r) = 0.0079 \, \frac{R_0^2 + R_c^2}{r^2 + R_c^2} \, M_\odot \text{pc}^{-3} \tag{4}$$

where $r$ is the galactocentric radius, $R_0 = 8.5$ kpc is the galactocentric distance of the Sun, and $R_c \approx 5$ kpc is the core radius, it is found that the optical depth towards the Large Magellanic Cloud (LMC) would be

$$\tau_{\text{LMC}} \sim 5 \times 10^{-7} \tag{5}$$

---

\* A similar calculation was carried out by Petrou (1981), but was not published.

if the halo is composed of compact objects (Paczynski 1986, Griest 1991). Although this optical depth is much lower than the fraction of intrinsic variable stars ($\sim 0.4\%$ for the LMC), microlensing events have many strong signatures which differ from all previously known types of variable star. For microlensing events involving a single point source, single lens and negligible accelerations, the events are symmetrical and achromatic, with a shape given by

$$A(t) = A(u(t)),$$
$$u(t) = \left[ u_{\min}^2 + \left( \frac{2(t - t_{\max})}{\hat{t}} \right)^2 \right]^{0.5}, \qquad (6)$$

where $A(u)$ is given by eq. 6, and $A_{\max} = A(u_{\min})$. Since the optical depth is so low, only one event should occur in any given star. If many events are found, additional statistical tests can be applied: the events should have a known distribution of peak amplifications (corresponding to a uniform distribution in $u_{\min}$); they should occur representatively across the HR diagram, and the event timescales and peak amplifications should be statistically independent.

Three groups have reported detections of candidate microlensing events. Our MACHO collaboration has reported three candidate events towards the LMC (Alcock *et al.* 1993, Alcock *et al.* 1995a) and four towards the galactic bulge (Alcock *et al.* 1995b); we have recently increased the bulge total to $\sim 45$ events (Bennett *et al.* 1995). The EROS collaboration has reported two events towards the LMC (Aubourg *et al.* 1993); and the OGLE collaboration has reported a total of 12 events towards the Galactic bulge (Udalski *et al.* 1993, 1994a). Also, three further groups have observations in progress; the DUO collaboration has $\sim 12$ preliminary candidates towards the bulge, and the AGAPE and VATT-Columbia groups have begun searches for microlensing towards M31.

In this paper, we present results from analysis of most of our first year's observations towards the LMC; this dataset consists of 9.5 million lightcurves of stars in the central 10 square degrees of the LMC, spanning approximately 400 days. In Section 2 we discuss the observations and photometric reductions; in Section 3 we discuss the automated search for microlensing events and the resulting candidates, and in Section 4 we provide an outline of the Monte-Carlo simulations used to measure our detection efficiencies. In Section 5 we compare our results with predicted event rates from simple models of the galactic halo and other populations, and we summarise our conclusions in Section 6.

## 2. Observations and Photometric Reductions

The MACHO project has full-time use of the 1.27-meter telescope at Mount Stromlo Observatory, Australia, for a period of at least 4 years from July 1992. The telescope was recommissioned especially for this project, and a computer-controlled pointing and drive system was installed. A system of corrective optics has been installed near the prime focus, giving a focal reduction to $f/3.9$ with a $1^o$ diameter field of view. A dichroic beamsplitter and filters provide simultaneous images in two passbands, a 'red' band (approx. 5900–7800 Å) and a 'blue' band (approx. 4500–5900 Å) Two very large CCD cameras are employed at the two foci; each contains a $2 \times 2$ mosaic of $2048 \times 2048$ pixel Loral CCD imagers. The pixel size is 15 $\mu m$ which corresponds to $0.63''$ on the sky, giving a sky coverage of $0.7 \times 0.7$ degrees. Each chip has two read-out amplifiers, and the images are read out through a 16-channel system and written into dual-ported memory in the data acquisition computer. One amplifier on one CCD in the red focal plane does not function. The readout time is 70 seconds per image, and the noise is $\sim 10$ electrons rms, with a gain of $\sim 1.9\,e^-/\mathrm{ADU}$; the images are written to disk and then saved on Exabyte tape. Details of the camera system are given by Stubbs *et al.* (1993) and Marshall *et al.* (1994).

Observations are obtained during all clear nights and part nights, except for occasional gaps for telescope maintenance. The default exposure times are 300 seconds for LMC images, 600 seconds for the SMC and 150 seconds for the bulge, so over 60 exposures are taken per clear night. As of 1995 April, over 30000 exposures have been taken with the system, of which about 19000 are of the LMC, 2000 of the SMC and 9000 of the bulge. The images are taken at standard sky positions, of which we have defined 82 in the LMC, 21 in the SMC and 94 in the bulge.[†]

Although the LMC is best placed for observing in the Southern summer, its high southern declination $\sim -70^o$ means that it is visible for at least part of the night during the entire year. The LMC is our primary target due to its sensitivity to halo microlensing; we therefore observe it at most times when it is above an elevation $\gtrsim 20^o$, and observe the SMC and Bulge otherwise. In the first year's observations, we concentrated on observing a subset of our LMC fields with good time resolution, to provide a useful test for substellar Machos in the mass range $10^{-4} - 10^{-1}\,\mathrm{M}_\odot$ which typically produce microlensing events of $\sim 1 - 30$ days duration (Griest 1991). Therefore, we defined 22 high priority fields which were observed twice per night when possible, while the remaining fields were observed approximately weekly to search for much longer events.

In the present paper, we consider **only** the data from these 22 well-sampled fields, located in the central $\sim 5^o \times 3^o$ of the LMC. The positions of these fields are are available via the WWW at the URL given before. The observations described here comprise 5169 images, covering a timespan of 409 days from 1992 July 21 to 1993 Sept 03. Some observations of the LMC were obtained on 242 of those nights. The mean number of exposures per field is $5169/22 = 235$, with a range from 140 to 350. This sampling varies quite substantially among our fields, since we usually observed the fields in a fixed order so that our "highest priority" fields were always observed even on partially clear nights, and were frequently observed twice per night. The number of monitored stars per field ranges from 616,000 to 265,000.

---

[†] Coordinates of the field centers are available on the WWW,
at URL: http://wwwmacho.anu.edu.au .

## 2.2 Photometric Reductions

Photometric measurements from these images are made with a special-purpose code known as SoDoPHOT (Bennett *et al.* 1995), derived from DoPHOT (Schechter *et al.* 1994). First, one image of each field with good seeing and dark sky is chosen as a 'template image'. This is processed in a manner similar to a standard DoPHOT reduction except that after one color of the image has been reduced, the coordinates of the stars found in the first color are used as starting points for the positions of stars in the second color, which improves the star matching between colors. (The final positions of the matched stars are forced to be the same in both colors, after allowing for differential refraction.) This procedure provides a 'template' catalog of stellar positions and magnitudes for each field.

All other images are processed in 'routine' mode, which proceeds as follows. First the image is divided into 120 'chunks' of $\sim 512 \times 512$ pixels, and for each chunk $\sim 30$ bright unsaturated stars are located and matched with the template. These stars are used to determine an analytic fit to the point spread function, a coordinate transformation, and a photometric zero point relative to the template. Then, all the template stars are subtracted from the image using the model PSF and coordinate transformation; noise is added to the variance estimate for each pixel to allow for errors in the subtraction. Next, photometric fitting is carried out for each star in descending order of brightness, by adding the analytic model of the star back to the subtracted frame and fitting a 2-parameter fit to the star's flux and sky background, with pixels weighted by inverse variance, while the model PSF and computed position of the star are kept fixed. When a star is found to vary significantly from its template magnitude, it and its neighbors undergo a second iteration of fitting. For each star, the estimated magnitude and error are determined, along with 6 other parameters (quality flags) measuring the object 'type' (single/blended etc.); the $\chi^2$ of the PSF fit; the 'crowding', i.e. the amount of flux contributed from nearby stars; the weighted fractions of the PSF masked due to bad pixels and cosmic rays respectively; and the fitted sky value. The photometric error estimate is the PSF fit uncertainty (as in DoPHOT) with a 1.4% systematic error added in quadrature. These routine reductions take approximately 1 hour per image on a Sparc-10 for a field with 500,000 stars in each color. The set of photometric datapoints for each field are re-arranged into a time-series for each star, combined with other relevant information including the seeing and sky brightness, and then passed to an automated analysis to search for variable stars and microlensing candidates.

## 3. Event Detection

We use a two–stage process to minimize the chances of mistaking either intrinsic stellar variability or systematic errors in photometry for microlensing. For the former, we exclude classes of stars that are prone to variability. For the latter, individual data points are required to meet certain acceptability criteria in order to be considered valid, using the various flags and fit parameters reported by the photometry. We determine the relationship between the quality flags and apparent 'bad' measurements as follows: we defined a set of 'non-variable' stars using a robust $\chi^2$ measure, which is designed to reject consistently variable stars while including stars with occasional discrepant data points. For these stars, we then examine the percentiles of the distribution of $\Delta m/\sigma$, where $\Delta m$ is the residual from the median magnitude and $\sigma$ is the error estimate. This is computed for many distinct bins of each flag. As expected, data points with large values of the various flags generally show a significant non-gaussian tail of outliers with large values of $\Delta m/\sigma$; thus, we set cuts on the various flags so as to reject most such outliers. We also mark measurements taken when the stellar image is within a few pixels of the edge of a CCD chip. Data points failing any of these cuts are marked as 'suspect'; they are retained in the database, but are not used in the microlensing or variability searches. These cuts reject approximately 15% of all data points.

We then define a set of 'useful' stars by requiring that each lightcurve have at least 7 simultaneous red-blue data points passing the above cuts, that it is classified as stellar and not very close to a chip boundary in the template image. Lightcurves failing any of these cuts are rejected from the microlensing search; this rejects some $\sim 15\%$ of all lightcurves representing mostly faint stars which are severely crowded, or are undetected in one colour in the template.

We also exclude stars redder than $V - R > 0.9$ from the microlensing search (but not from the variable search). This removes the reddest 0.5% of stars from the microlensing search. These stars are often long-period variables which almost always trigger the level–1 criteria and would dominate the total number of triggers if included.

The microlensing search through the light curve database proceeds in three stages: first, the time-series are convolved with a set of filters of durations 7, 15 and 45 days in order to search for peaks of any kind. Any lightcurve with a significant peak in both colors in any filter is tagged as a 'level–1' trigger; $\sim 0.5\%$ of the lightcurves pass this trigger. For these level–1 lightcurves a 5-parameter fit to a microlensing event is made, where the parameters are the un-amplified red and blue fluxes $f_{B0}, f_{R0}$, the peak amplification $A_{\max}$, the time of peak amplification $t_{\max}$ and the Einstein-diameter crossing time $\hat{t}$. Thus, the fitted flux of the star in each colour is given by

$$f_B(t) = f_{B0} A(u(t)) \tag{7}$$
$$f_R(t) = f_{R0} A(u(t)),$$

where $A(u)$ is given by eq. 6, and $A_{\max} = A(u_{\min})$. Then, a set of statistics describing significance level, goodness of fit, achromaticity, crowding, temporal coverage of the event, etc. are calculated. Events above a modest significance level are tagged as level–1.5 events and are output as ASCII files, along with their associated statistics; these level–1.5 candidates are then subjected to more rigorous selection criteria, which may be easily modified, to search for final 'level–2' microlensing candidates.

Figure 1 summarizes the 'level–2' cuts we have used for this analysis. These cuts include:

1) We make a cut on the SoDOPHOT crowding parameter which uses the noise array (defined as in DOPHOT, Schechter *et al.*, 1994) to indicate stars whose central pixel may get more flux from neighboring stars than from themselves. Stars with very close, brighter neighbors are removed with this cut.

2) We require that the entire 'peak region' of the light curve (*i.e.* the part with $A > 1.1$) is contained within the timespan of observations, and that there are at least 3 data points (in either colour) on both the rising and falling portions of the light curve, i.e. the portions between $A = 1.1$ and the peak.

3) We define $\chi^2_{\rm ml}$(d.o.f.) to be the overall $\chi^2$ per d.o.f of the whole lightcurve, and $\chi^2_{\rm peak}$(d.o.f.) to be the $\chi^2$ in the interval where the fitted $A > 1.1$; we require $\chi^2_{\rm ml}$(d.o.f.) $< 3$ and $\chi^2_{\rm peak}$(d.o.f.) $< 4$.

4) We demand that the color of the star in the peak region be consistent with the median color formally at the 99.7% confidence level and that each event includes at least 8 data points which are $> 1 - \sigma$ above the median brightness.

5) We require that $A_{\max} > 1 + 2\sigma_f$ where $\sigma_f$ is the average estimated flux error, in units of the median flux.

6) We remove long timescale events, where we do not have a good measurement of the unamplified light curve. We require $\hat{t} < 250$ days, and the FWHM of the microlensing fit $< 100$ days.

7) Among the most important cuts is on the improvement in absolute $\chi^2$ of the microlensing fit relative to a constant flux fit: we define

$$\Delta\chi^2 = (\chi^2_{\rm ml} - \chi^2_{\rm const})/\chi^2_{\rm ml}({\rm d.o.f.}),$$

and we require $\Delta\chi^2 > 200$.

8) We demand that the fitted $A_{\max} > 1.5$.

9) Although not shown in Figure 1, we remove a particular 0.03 square degree area of the sky (9a) and all stars with $V < 17.5$ (9b) in order to ensure that particular types of variable stars are excluded by at least two distinct cuts. The characteristics of these variables are discussed in the next section.

The 'final' cuts for lightcurves passing criteria (1)-(5) are shown in Figure 2. The events indicated by filled circles are the ones which pass cuts (6) and (9), while lightcurves which fail cut (6) are indicated by open circles. The crosses and open triangles represent lightcurves of events which are related to our most important variable star backgrounds which are discussed in some detail below. All of these events fail at least 2 different cuts. The crosses indicate lightcurves which fail our $V > 17.5$ cut. Most of these lightcurves have $A_{\max} < 1.5$. The open triangles indicate stars from a very small region of the sky which may contain some very young stars. Their lightcurves all fail cut (6) as well as the explicit cut (9a)

The most significant event which fails the timescale cut (6), does not fall in the small region referred to in cut (9). It is a very faint star ($V > 22$) with a best fit $A_{\max} = 20$ and $\Delta\chi^2 \approx 4000$ which is not clearly detected when it is not amplified. Recent observations of this star have indicated two subsequent brightenings of similar amplitude. It may be an 'old nova' as discussed by Della Valle (1994).

The large number of events at low $\Delta\chi^2$ generally contain small bumps attributable to low-level, systematic photometry errors. A majority of the low level bumps appear near day 98. On day 22, it was noted by the observer while focussing that there was a marked increase in the astigmatism of the system. It was discovered that a primary mirror radial support pad had broken loose. The pad was reglued and the telescope realigned on day 26. Some time after this, the pad apparently began coming loose again until it was discovered on day 98. The telescope was aligned again and the pad resecured. At this point in time, the light curve of the "alignment" bumps all returned to baseline. Visual inspection of data frames from a few of the fields showed that images undergoing these alignment bumps were sources with close neighbors of roughly similar brightness at about the same position angle as the astigmatism induced image elongation. Figure 3 shows a representative light curve of one of these "alignment" bump events. (The events with large $A_{\max}$ and low $\Delta\chi^2$ generally have the fitted peak falling in a data gap, so the significance is low).

### 3.1 VARIABLE STAR BACKGROUNDS

The events with highly significant $\Delta\chi^2$ but low $A_{\max}$ occur in the light curves of bright main sequence stars with $V \sim 15 - 17$ which are generally at constant brightness, but show occasional brightening by about 10-30%. We will refer to these stars as "bumpers". These stars most often brighten more in our red passband ($5900-7800\text{Å}$) than our blue passband ($4500-5900\text{Å}$). For more than half of these stars the episodes are clearly asymmetric and shorter than about 50 days with a more rapid brightening than dimming (see Figure 4 top panel). The rest show more symmetric, and longer episodes (see Figure 4 lower panel). The fit to microlensing can be relatively good for the long events, and also for the short events if they are not well sampled. We have obtained spectra of seven of these stars and two of them showed strong H$\alpha$ and H$\beta$ emission. All of them showed evidence for filled cores in the Balmer lines. These stars are probably similar to, or related to, galactic Be stars. Percy *et al.* (1988) and Percy & Attard (1992) have noted possible outbursts of a similar nature in galactic Be stars. All of the bumpers identified in our first year's data have shown additional significant photometric deviations in our subsequent years' data. A complete discussion of this phenomenon will be found in Alcock *et al.* 1995 (in preparation). The frequency of these low-amplitude events in bright, main sequence stars, their poor fit to microlensing, their chromatic nature and the high frequency of repetition show that these are not microlensing events. Because of our current lack of understanding of this phenomenon and its progenitor stars, we choose to eliminate all stars brighter than V=17.5 and to require $A_{\max} > 1.5$ for consideration of an event as microlensing.

There is another class of variable stars which appear similar to microlensing events; all of these are found in field-82 in a region about 5 arc minutes in diameter centered about 0.7° West of 30 Doradus. As seen in Figure 5, this region has a great deal of nebulosity. (There is perhaps one other region in the 11 square degrees surveyed where there is as much nebular emission.) Our microlensing search has yielded a large number of microlensing triggers in this region, and there are 14 lightcurves which pass all our cuts except for the cuts on the timescales of the events. The best fit microlensing light curves for these lightcurves yield $\hat{t}$ values ranging from 190 to 550 days, and the best fit peak amplifications range from 1.6 to 10. Light curves of two of these events are shown in Figure 6. Many of these 14 stars appear to be quite close to local maxima in the nebulosity, suggesting that they are related to T Tauri stars and that they are located in a large star-forming region.

A possible alternative interpretation, however, is that MACHOs could occur in clusters (e.g. Maoz 1994, Wasserman and Salpeter 1994, Moore and Silk, 1995) so that microlensing event positions might be highly correlated on the sky. The very long time scales of the events might be taken to imply that there is a MACHO cluster with a very small transverse velocity on this line of sight. Of course, it is unlikely that such a cluster would be aligned with a region of such high nebulosity. A more severe difficulty with the MACHO-cluster explanation is that all 14 of these stars are near the main sequence and fainter than $R = 19.5$ which puts them into the faintest two thirds of the stars in this field. Because of the magnitude dependence of our efficiencies, we expect that most of our detected microlensing events will occur in the brighter stars with $R < 19.5$. The probability that 14 lensing events would all occur in stars fainter than our expected median is $2^{-14} = 6 \times 10^{-5}$. Thus, it is clear that these are not genuine microlensing events. In the current analysis, we imposed the cuts $\hat{t} < 250$ days and event FWHM $< 100$ days to reject these lightcurves. We have also excluded a region of about 0.03 square degrees of the sky containing these stars, to ensure that these objects are rejected by two distinct cuts.

The process of establishing an appropriate set of selection criteria has been subjective: the various processes that might mimic microlensing were not known prior to the start of the project. We believe that we have arrived at a reasonable set of selection criteria and that microlensing has in fact been detected, but there is no guarantee that *every* candidate event that passes the selection criteria is genuine microlensing.

### 3.2 MICROLENSING CANDIDATES

Four database objects pass all the cuts described above;[†] their light-curves are shown in Figure 7, their positions and fit parameters are listed in Table 1, and finding charts are available upon request and via WWW. Two of these four objects correspond to a single star falling in an overlap between two of our fields; our observations of this star from **one** field were shown in Alcock *et al.* (1993). The two fields are treated completely independently in our reductions, with different templates and measurements derived from distinct CCD frames, so the good agreement between these two sets of results is a useful check of our reliability. We refer to the separate lightcurves as events 1a and 1b. (This event has also been confirmed in the independent data of Aubourg *et al.* 1993).

As discussed in Alcock *et al.* (1993), there are many characteristics of event 1 which seem to argue strongly that this event is due to microlensing. The peak is symmetrical and achromatic to good accuracy; no variations are detectable in the remainder of the light-curve; and its shape is consistent with microlensing, with the broad wings and narrow peak which are characteristic of high-amplification events. There is a minor fit discrepancy near the peak, with the point just before the peak being high in both colours and the following point being low in both colours. This has been suggested as indicating that the lens is a binary system (Dominik & Hirshfeld 1994). However, we caution that if the latter point is arbitrarily eliminated, a good fit with $A_{\max} \approx 8$ can be made through the former point, so the binary interpretation involves 3 additional free parameters to fit effectively a single discrepant observation. Unfortunately, the field containing event 1b does not have data on the corresponding nights, so does not resolve this question.

---

[†] A recent re-analysis of the data discussed in this paper reveals a very high magnification event ($A_{max} \sim 50$) in a faint star that is not included in the analysis described here, because it did not pass cuts (1) and (4).

| Event | RA (2000) | Dec (2000) | V | V-R | $t_{\max}$ (days) | $\hat{t}$ (days) | $A_{\max}$ | $\chi^2$ |
|---|---|---|---|---|---|---|---|---|
| 1 | 05 14 44.5 | -68 48 00 | 19.6 | 0.6 | $57.16 \pm 0.02$ | $34.8 \pm 0.2$ | $7.20 \pm 0.09$ | 1.34 |
| 2 | 05 22 57.0 | -70 33 14 | 20.7 | 0.4 | $121.62 \pm 0.3$ | $19.8 \pm 1.3$ | $1.99 \pm 0.06$ | 1.41 |
| 3 | 05 29 37.4 | -70 06 01 | 19.4 | 0.3 | $154.8 \pm 0.9$ | $28.2 \pm 1.7$ | $1.52 \pm 0.03$ | 1.01 |

**Table 1:** Parameters of the events. Columns 4 & 5 show approximate magnitude and color of the lensed stars. Typical uncertainties are ~0.1 mag. Columns 6–8 show the parameters of the best-fit microlensing models: time of peak amplification (Julian days–2449000), the event duration $\hat{t}$, and the peak amplification factor, with the *formal* one sigma errors (derived from the covariance matrix of the fit). Column 9 is the $\chi^2$ per degree of freedom for the microlensing fit.

The spectrum of this star is consistent with a normal late F/early G-type giant with the radial velocity of the LMC (della Valle 1994). No variations have been found during our continued monitoring of this star from 1993 August - 1994 September.

Clearly, the two low-amplitude candidates (events 2 & 3) are of very much lower significance than event 1; this is due both to the lower amplifications and the fact that event 2 occurs in a considerably fainter star. Therefore, it is not possible to make a strong claim that these events are actually microlensing, though they are certainly consistent with microlensing events. It is unlikely that events 2 & 3 are due to observational errors, since they are reasonably well separated from the 'noise' events in the $\Delta\chi^2$ plot. Furthermore, event 2 is also located in a field overlap, and shows similar brightening in the data from the second field (though it does not pass the selection cuts in the second field due to its closeness to a boundary).

We have obtained a spectrum of the star involved in Event 3, which shows no obvious peculiarities; a spectrum of the Event 2 star has been obtained by M. Della Valle (private communication), and also appears normal, though the signal-to-noise is low due to the faintness of the star.

The peak amplifications of the three events correspond to dimensionless impact parameters $u_{\min} = 0.14, 0.55$ and $0.82$, while the predicted distribution should be a uniform distribution modified by our detection efficiency. Using the detection efficiencies described in the next Section, we find a K-S probability of 0.36 that these three events come from an efficiency corrected theoretical microlensing distribution. This is clearly consistent with the microlensing hypothesis, but with only three events it not a powerful test of microlensing. We note that for 43 events towards the Galactic bulge, we have found good consistency with the predicted distribution.

Thus, we conclude that it is highly probable that event 1 was caused by microlensing rather than intrinsic variation of the star, while events 2 & 3 are consistent with microlensing, though intrinsic stellar variability remains a possibility.[†] In later Sections, we consider a range of implications derived from 1–3 events interpreted as microlensing.

## 4. Detection Efficiency

Based on the halo model of eq. 4, if the dark halo were made entirely of compact objects of unique mass $m$, the rate of microlensing events with $A > 1.34$ would be $\Gamma = 1.6 \times 10^{-6} \sqrt{M_\odot/m}$ events/star/yr. Thus if our experiment were 100% efficient, our 'exposure' of $E = 9.7 \times 10^6$ star yr would lead to an expected number of events $N \approx 16 \sqrt{M_\odot/m}$ in the present dataset, i.e. $N > 50$ for substellar Machos; but clearly, a reliable estimate of our detection efficiency is essential before we can draw conclusions about the abundance of Machos in the halo. We define our detection efficiency $\mathcal{E}(\hat{t})$ as the fraction of events in all monitored stars with timescale $\hat{t}$, amplification $A_{max} > 1.34$, and $t_{max}$ within our sampling period, which would be accepted by our standard selection criteria. Note that since one of our cuts is $A_{max} > 1.5$, our efficiency thus defined can never go above 83%.

Due to bad weather, variations in seeing and lunar phase, and the changing hour angle of the LMC, our observations have irregular spacing, and the photometric errors vary considerably both from star to star and from image to image. There is an additional source of inefficiency arising from stellar crowding, due to the fact that many of the 'stars' we monitor may consist of unresolved blends of two or more individual stars. Thus our detection efficiency cannot be computed analytically, and must be estimated via Monte-Carlo simulations.

### 4.1 SAMPLING EFFICIENCY

As an initial estimate, we have carried out a relatively simple Monte-Carlo calculation to estimate our microlensing detection efficiency under the assumption that all our catalogued stars are in fact point sources, and that the photometry code recovers all excess flux in an unbiased way. (We drop these assumptions in § 4.2). The distribution of microlensing amplifications is model-independent and given theoretically, and the expected distribution of peak times will be uniform, apart from a very small seasonal modulation (Griest 1991) due to the Earth's orbit, which we ignore. Hence, for comparison with models we only require the efficiency as a function of event timescale $\mathcal{E}(\hat{t})$. [For verifying consistency with the theoretical $A_{max}$ distribution, clearly we require $\mathcal{E}(A_{max}, \hat{t})$; but given 3 events, this test is rather weak, as seen in the previous section.]

The reduced photometry database used in this paper occupies an unwieldy 50 GB of data, so we first constructed a manageable subsample by extracting $\sim 1\%$ of all lightcurves at random from the database, i.e. 92,000 lightcurves including field overlaps. We added a single simulated microlensing event to each star, as follows: we generated event parameters at random with the time of peak amplification $t_{max}$ uniformly distributed over a 415-day interval slightly wider than the data timespan, the minimum impact parameter $u_{\min}$ uniform with $0 \le u_{\min} \le 1.2$, and the timescale uniform in $\log(\hat{t})$ with $0 \le \log_{10}(\hat{t}/\text{days}) \le 2.5$. For each observation in the simulated microlensing event, we generated a simulated flux $f_{sim} = f_{obs} + (A-1)f_{med}$, where $A(t)$ is the theoretical amplification at the time of the real observation, $f_{obs}$ is the observed flux in this observation, and $f_{med}$ is the median observed flux over all observations for this star. This procedure increases the flux by the theoretical amount while preserving the real scatter inherent in the measurements. We increased the estimated error of the real data point by a factor $\sqrt{A}$ (in flux units), i.e. decreased it by $1/\sqrt{A}$ in magnitude units.

---

[†] The EROS group has informed us that they have some evidence, at very low signal to noise, that the star we report as event 2 underwent a brightening event about one year prior to our start of data taking (Milsztajn, private communication). We are continuing to monitor this star, and so far have seen no evidence for variation other than the magnification discussed here.

We then ran the standard microlensing search software on the resulting simulated dataset and matched up the input and output microlensing fit parameters for those events passing the selection criteria; the agreement is generally very good, as shown in Figure 8a.

The resulting detection efficiency $\mathcal{E}_1(\hat{t})$ is the number of recovered events relative to the number input per $\hat{t}$ bin with $A_{\max} > 1.34$. This is shown in Figure 9; it shows a broad peak at $\mathcal{E} \sim 0.3$ for timescales $20 \lesssim \hat{t} \lesssim 120$ days, and declines steeply for events outside this range; though it remains above 0.1 for events as short as 6 days. The decline for short events is as expected due to weather gaps, our constraint that at least 8 data points (i.e. 4 observations) during an event are required, and our use of the $\Delta\chi^2$ criterion which (other factors being equal) scales proportional to the number of data points during an event. The decline for long events is due to our explicit cuts that $\hat{t} < 250$ days and the event FWHM $< 100$ days; these are necessary as mentioned before to reject long-timescale variables.

The peak efficiency of $\mathcal{E} \approx 0.35$ may at first glance seem low; however, we have visually inspected a large number of simulated events both passing and failing the cuts, and find that 'good-looking' events are rarely missed by the software. The main reasons for non-detection of events are as follows:

i) We have an explicit efficiency factor of 0.83 due to our criterion $A_{\max} > 1.5$, rather than the 'canonical' $A_{\max} > 1.34$. This of course factors out in our conclusions, but it is convenient to define all efficiencies relative to this value since this facilitates comparison between different subsamples and different experiments.

ii) One half of one of our 4 CCD chips in the red camera is dead. Since the telescope is on an asymmetrical "German" mount, the focal plane can be rotated by $0°$ or $180°$ relative to the sky depending on the hour angle. The two possible orientations of the dead area mean that 1/4 of our stars have only half as frequent coverage in the red data, which varies seasonally; we require some coverage in both colours in the selection cuts, so this will reject up to 12.5% of events.

iii) Some of our stars ($\approx 15\%$) are completely rejected from the lensing search due to severe crowding with a brighter neighbour, suspect photometry flags in the template reduction, or non-detection in one colour in the template. Many more are so faint, with typical error bars $\sim 30\%$, that only high-amplification events with $A_{\max} \gtrsim 3$ in these lightcurves would pass our acceptance cuts.

iv) Some of our fields did not commence observations until 1993 October, and there are several gaps of $\sim 2$ weeks duration caused by telescope maintenance and storm systems; events peaking in these gaps are rarely detected.

The combination of these factors essentially accounts for the apparent moderate efficiency; each of these contributes a factor of $0.8 - 0.9$.

In the following section we discuss a refinement of this simple estimate, which accounts both for the blending of stars and for any biases which may exist in the photometry code. This is found to provide a significant but not severe reduction in efficiency relative to the simple estimate given above.

4.2 BLENDING EFFICIENCY

An accurate efficiency estimate is more difficult than is apparent at first sight, since there is an additional complication caused by the high degree of crowding in our fields. Due to this, many of the 'stars' in our database, especially faint ones, may actually be unresolved or marginally resolved blends of two or more fainter stars. This may occur either if two stars form a physical binary, or by chance superposition. Although the fraction of stars in binaries is thought to be over 50%, this only affects the lensing if the two components are of roughly equal luminosity, and the effect is expected to be relatively small (Griest & Hu 1992); thus we neglect this effect. However, due to the relatively poor seeing in our images, and the huge stellar density (over 1.2 million measured stars deg$^{-2}$ in our densest fields), the fraction of chance superpositions will not be negligible.

If one component of a blend comprising a fraction $f$ of the flux is microlensed, unless the star is a binary with separation $\lesssim r_E/x$ its neighbours will not be lensed, so the apparent amplification of the blend will naively be

$$A_{obs} = (A_{true} - 1)f + 1 < A_{true}, \qquad (8)$$

where $A_{true}$ and $A_{obs}$ are the true and observed amplifications. The real effect is complicated by the fact that $f$ typically varies with observing conditions. The position of a fitted 'star' is fixed in our photometry code to its template position, effectively the centroid of the blend, however the positions of individual components may differ from this by up to $\sim 1''$. The relative flux contribution to the blend from various components depends not only on their intrinsic flux but on their displacement from the fit center in units of the variable seeing FWHM.

Of course, we cannot tell which individual stars are actually blended; but we can model this effect statistically by adding a large number of artificial stars into a sample of our data frames, and re-running the photometry code to compare the input and recovered fluxes. We have therefore selected two $5' \times 5'$ sub-regions of a single field for this experiment. This field is on the edge of the LMC bar and the two regions, on opposite edges of the field, differ in stellar density by almost a factor of 2. We have extracted 20 two-color images of this field which span a wide range of seeing and sky brightness. For each of these two sub-regions in both colors, and each of the 20 observations, we have measured an empirical point spread function from $\sim 20$ bright stars using DAOphotII (Stetson 1992).

Then, 196 artificial stars have been added on a spatial grid at a range of amplifications using the empirical PSFs; each artificial star has been added in at 12 different amplifications from $0.9\times$ to $26\times$, and the photometry code has been run on all these. The artificial stars have 'un-amplified' luminosities drawn at random from our observed luminosity function. Photometric response functions are obtained by simply associating the input artificial star amplification with the photometry of the nearest detected object. The entire procedure has been repeated 30 times with different artificial star positions and luminosities, to produce a catalogue of 2 regions × 196 stars × 30 runs × 20 frames × 12 amplifications = 2.8 million artificial data-points.

This dataset is then fed into another Monte-Carlo simulation, similar to that in the previous Section; but now, instead of adding the 'theoretical' additional flux to each data-point, we use a measured 'response' from the artificial star database as follows: for each observed star, we randomly generate a set of microlensing event parameters as before. We 'associate' an artificial star with this real star, by choosing a near neighbour in a (magnitude, crowding) plane, with an additional 'usage' criterion to force reasonably uniform sampling of the set of artificial stars.

For each real data point on the observed star, we choose the frame in the artificial sample which best matches the seeing and sky brightness of the real observation; this gives a look-up table of 'recovered' photometry vs 'input' amplification for the 'associated' artificial star in the appropriate observing conditions. An artificial 'data-point' is obtained by interpolating between the two nearest input amplifications in this table. The simulated flux is then given by $f_{\rm sim} = f_{\rm int} + f_{\rm obs} - f_{\rm med}$, where $f_{\rm int}$ is the interpolated flux, $f_{\rm obs}$ is the flux of the real star and $f_{\rm med}$ is the median or baseline flux of the real star. The other parameters such as the flux error and the flags are generated in a corresponding way. This procedure retains the 'real' measurements when the simulated amplification is small, and is close to the purely 'artificial' measurements when it becomes large.

The process is repeated several times for all the 1% subsample of real lightcurves, with the standard analysis run as before, to obtain a revised efficiency function which we call $\mathcal{E}_2(\widehat{t}, m)$. This efficiency is tabulated as a function of *input* stellar magnitude, not of the recovered or blended magnitude. It is therefore an underestimate of our efficiency for detecting microlensing of a blended object, since a blend consists of two or more components, each of which may be lensed. Thus $\mathcal{E}_2(\widehat{t}, m)$ must be integrated over a stellar luminosity function to yield the true experimental exposure $E$.

Shown in Figure 9 is $\mathcal{E}_2(\widehat{t})$ obtained by integrating $\mathcal{E}_2(\widehat{t}, m)$ over several different luminosity functions shown in Figure 10 and normalised to the number of detected objects. A firm lower bound, as mentioned above, is $\mathcal{E}_2$ integrated over the observed luminosity function. More physical approximations are obtained by using two estimates of the true stellar luminosity functions. The first of these is the luminosity function of a relatively uncrowded field outside of the LMC bar, explicitly illustrating the effect of crowding. The second is an extrapolation of the luminosity function beyond the clump giants using a power law $\propto 10^{0.5m}$, roughly that of the main sequence obtained by Elson *et al.* (1994) for V $\lesssim$ 22.5. In all cases we have excluded contributions to the exposure by stars fainter than 21.5 as our efficiency is very poorly determined beyond this. There is likely to be a small but measurable contribution to the exposure from these faint stars so the blend efficiency is an underestimate in this sense. Finally, our upper bound is the time sampling efficiency from the previous section.

In addition to reducing detection efficiency, the effects of crowding also give rise to a bias in the microlensing fit parameters: the measured amplification is systematically suppressed and the measured duration is systematically shortened. (This latter effect is not intuitively obvious, but is due to the fact that lower amplification events are less sharply peaked and have a broader FWHM for a given $\hat{t}$). Figure 8 shows equal likelihood contours for recovering input microlensing fit parameters of artificial events. Figure 8a shows the contours for $\sim 1.8 \times 10^5$ events created without crowding effects as in §4.1, *i.e.* single star events, and shows that the parameters are accurately recovered. By contrast, Figure 8b shows the pronounced bias toward underestimating $\hat{t}$ and $A_{\max}$ for $\sim 6 \times 10^5$ events created with crowding effects. We take into account this effect in the analysis of Section 5 by correcting the measured $\hat{t}$ of our events by the factor $<\hat{t}_{real}>/<\hat{t}_{measured}>$ derived from many artificial events of similar $\hat{t}$ on stars of similar brightness.

To summarise this section, the real experimental efficiency may differ from our best estimate by at most $\approx 20\%$; this does not seriously compromise the conclusions in the following Sections.

## 5. Predicted Event Rates

Although the optical depth $\tau$ for microlensing is the most commonly discussed quantity, the rate of events $\Gamma$ is more important for comparison with observation, since the observed number of events is given by Poisson statistics.

The total rate $\Gamma$ is related to the optical depth $\tau$ and mean timescale $\langle \hat{t} \rangle$ by

$$\Gamma = \frac{4}{\pi} \frac{\tau}{\langle \hat{t} \rangle}. \qquad (9)$$

The factor of $4/\pi$ arises because for a given event, $\hat{t}$ is the time for the lens to traverse a distance of $2r_E$ relative to the line-of-sight. The duration for which the lens is within $b < r_E$ of the line-of-sight ($A > 1.34$) is $t_e = \hat{t}\sqrt{1 - u_{\min}^2}$. Since the distribution of events in $u_{\min}$ is uniform, $\langle t_e \rangle = \frac{\pi}{4}\langle \hat{t} \rangle$, and it is clear that $\Gamma \langle t_e \rangle = \tau$.

In the next subsections, we first estimate event rates for the 'standard' halo of Griest (1991), and then for a wide set of realistic halo models, and compare these with the observed event rate to set limits on the baryonic content of the halo. We then consider a likelihood-based method which uses the additional information given in the event timescales to find estimates of the Macho contribution to the halo, under the assumption that all three events are due to microlensing of halo objects.

### 5.1 Event Rates from the Standard Halo

As a baseline for comparison, we adopt the "standard" halo model of eq. 4, hereafter model S. This halo has a mass interior to the Solar radius of $3.2 \times 10^{10} \, M_\odot$, and a halo mass interior to 50 kpc of $4.1 \times 10^{11} \, M_\odot$; it predicts an optical depth to the LMC of $\tau = 4.7 \times 10^{-7}$. (This is slightly lower than the value $5.1 \times 10^{-7}$ given by Griest (1991) since we adopt an LMC distance of 50 kpc rather than 55 kpc.) For comparison, the halo models suggested by e.g. Caldwell & Ostriker (1981) and Bahcall *et al.* (1983) produce optical depths around $25-50\%$ higher than this value (de Rujula *et al.* 1994) so this is a moderately conservative choice.

For computing event rates, we need also to model the velocity distribution and mass function of the lenses. We use a delta function mass function of arbitrary mass $m$, and a Maxwellian velocity distribution with a 3-D velocity dispersion $\sigma_v = 270 \,\mathrm{km/s}$, as in Griest (1991). Then the theoretical event rate as a function of timescale $d\Gamma/d\hat{t}$ is

$$\frac{d\Gamma}{d\hat{t}} = \frac{32 L}{\hat{t}^4 m v_c^2} \int_0^1 \rho(x) r_E^4(x) \exp\left(-\frac{4 r_E^2(x)}{\hat{t}^2 v_c^2}\right) dx, \qquad (10)$$

where the complete formulas, a description of the variables, and a discussion of the approximations made, are given in Appendix A.

The theoretical event rate for halo model S, with a delta-function Macho mass function at $1\,\mathrm{M}_\odot$ is shown in Figure 11; for other masses one simply scales the x-axis by $\sqrt{m}$ and the y-axis by $m^{-1}$, hence the total event rate scales as $m^{-0.5}$. The total event rate is $\Gamma = 1.6 \times 10^{-6} \sqrt{\mathrm{M}_\odot/m}$ events/star/yr, and the mean timescale $\langle \hat{t} \rangle = 130 \sqrt{m/\mathrm{M}_\odot}$ days.

Given the efficiency estimates from the previous section, it is quite straightforward to obtain predictions of expected observational event rates for any halo model. For an experiment with total exposure $E$ and efficiency $\mathcal{E}$, the observed number of events will follow a Poisson distribution with a mean of

$$N_{\exp} = E \int_0^\infty \frac{d\Gamma}{d\hat{t}} \mathcal{E}(\hat{t}) \, d\hat{t}. \qquad (11)$$

For convenience, we define $\tilde{N}(m)$ to be the expected number of detected events for a halo composed entirely of Machos with unique mass $m$, for a given halo model. This function $\tilde{N}(m)$ for model S is shown in Figure 12, using our 'best' efficiency model from the previous Section; it peaks at $\tilde{N} \sim 25$ for $m \sim 0.01\,\mathrm{M}_\odot$ and falls off for higher and lower masses. There are two competing effects; for masses $m \sim 0.1 - 1\,\mathrm{M}_\odot$, most events have timescales of $\hat{t} \sim 30 - 100$ days, near the peak of our efficiency curve, and the fall in $\tilde{N}$ is due to the $m^{-0.5}$ factor in the event rate; but for $m \lesssim 10^{-3}\,\mathrm{M}_\odot$, most events are shorter than $\hat{t} \sim 3$ days, where our efficiency is very low, causing $\tilde{N}_{\exp}$ to fall for low masses.

We have seen that 3 events are observed with our selection criteria; we can use Poisson statistics and the fact that no more than 3 events were observed to set an upper limit of $N_{\exp} < 7.7$ at 95% confidence, **irrespective** of whether or not these 3 events were due to halo microlensing. * This immediately translates into an upper limit on the Macho fraction of the halo as a function of mass, $f_{\lim}(m) = 7.7/\tilde{N}(m)$, shown in Figure 13.

Of course, it is unlikely that a real halo would have a delta-function mass distribution, but we emphasise that upper limits on the Macho fraction of the halo for delta-function models over some range can also be applied to **any mass function** which is contained within the same range, as follows. Suppose that $\tilde{N}(m) \geq N_0$ for some constant $N_0$ for some mass range $m_1 \leq m \leq m_2$.

---

* If Machos occur in large clusters, the statistics may not exactly follow a Poisson distribution. However, from disk-heating arguments, the mass of the clusters must be $\lesssim 10^6\,\mathrm{M}_\odot$ (Lacey & Ostriker 1985; but see Wasserman and Salpeter 1994); hence there must be $\gtrsim 100$ clusters contained in the 11deg$^2$ solid angle observed here, and Poisson statistics remain valid for small numbers of events.

Define $\psi(m)\,dm$ to be the fraction of halo mass in Machos in the mass range $m, m+dm$, and $f$ to be the total fraction in Machos with $m_1 < m < m_2$. Then the expected number of events is

$$\begin{aligned}N_{\exp} &= \int_0^\infty \psi(m)\,\tilde{N}(m)\,dm \\ &\geq \int_{m_1}^{m_2} \psi(m)\,\tilde{N}(m)\,dm \\ &\geq N_0 \int_{m_1}^{m_2} \psi(m)\,dm = N_0 f.\end{aligned} \quad (12)$$

From Figure 12, we find that $\tilde{N}(m) > 7.7$ for the range $8 \times 10^{-5}\,\mathrm{M_\odot} < m < 0.3\,\mathrm{M_\odot}$; thus Machos in this range comprise $< 100\%$ of our standard halo at 95% confidence. Similarly, $\tilde{N}(m) > 15.4$ for the range $3 \times 10^{-4}\,\mathrm{M_\odot} < m < 0.06\,\mathrm{M_\odot}$, thus Machos in this range comprise less than 50% of the mass of the standard halo.

### 5.2 Model-Independent Limits on Machos

In §5.1 we placed strong constraints on our "standard" halo model S, which are valid whether or not our 3 events were caused by microlensing of halo objects. However, there are very few tracers of the dynamics of the outer galaxy, so there is considerable uncertainty in the halo parameters. Thus it is important to examine a range of halo models before drawing conclusions about the Macho content of the halo. In this section we examine a wide range of halo models in an attempt to find more model independent limits on the baryonic content of the Milky Way Halo.

We restrict ourselves to the simple model 'S' of eq. 4, and to the "power-law" models of Evans (1993; 1994). The power-law halos are self-consistent models with analytic velocity distributions, convenient for computing event rates $d\Gamma/d\hat{t}$ (Alcock *et al.* 1995c; Evans & Jijina 1994). This set of models is specified by a normalization velocity $v_a$, the halo flattening, $q$, the asymptotic slope of the rotation curve, $\beta$, as well as $R_c$ and $R_0$. The power-law halos have a density profile

$$\rho = \frac{v_a^2 R_c^\beta}{4\pi G q^2} \frac{R_c^2(1+2q^2) + R^2(1-\beta q^2) + z^2(2-(1+\beta)q^{-2})}{(R_c^2 + R^2 + z^2 q^{-2})^{(\beta+4)/2}}, \quad (13)$$

where $R^2 = r^2 + z^2$, and $z$ is the distance above the plane of the disk. Here $q = 1$ corresponds to a spherical halo and $q = .7$ corresponds roughly to an E6 halo, while $\beta = 0$ gives an asymptotically flat rotation curve, $\beta < 0$ gives a rising rotation curve and $\beta >$ gives a falling rotation curve. The distribution function gives an isotropic velocity distribution and can be found in Alcock *et al.* (1995c).

In normalizing the dark halo, the size and shape of the stellar disk is important, since some (or even much) of the local centrifugal balance is given by the disk mass. We model the disk as a thin exponential disk specified by a scale length, $R_d \approx 3.5$ kpc, and a local column density $\Sigma_0$. We consider $\Sigma_0$ in a range from the canonical value of $50\,\mathrm{M_\odot pc^{-2}}$ to the extreme "maximal disk" value of $100\,\mathrm{M_\odot pc^{-2}}$. After specifying the disk, the normalization velocity $v_a$ is set by requiring a total rotation speed at $R_0$ of within 15% of the I.A.U. value $v_c \approx 220$ km/sec. Note, in adding the disk contribution we have sacrificed the self-consistency of the power-law model, but for LMC microlensing this should not be a large effect (Evans & Jijina 1994). For more description of these models, see Evans (1994); formulae for the microlensing event rate, optical depth, and event duration distribution are given by Alcock *et al.* (1995c).

| Model | S | A | B | C | D | E | F | G |
|---|---|---|---|---|---|---|---|---|
| Description | med. | med. | large | small | E6 | max disk | big disk | big disk |
| $\beta$ | - | 0 | -0.2 | 0.2 | 0 | 0 | 0 | 0 |
| $q$ | - | 1 | 1 | 1 | 0.71 | 1 | 1 | 1 |
| $v_a$ (km/s) | - | 200 | 200 | 180 | 200 | 90 | 150 | 180 |
| $R_c$ (kpc) | 5 | 5 | 5 | 5 | 5 | 20 | 25 | 20 |
| $R_0$ (kpc) | 8.5 | 8.5 | 8.5 | 8.5 | 8.5 | 7.0 | 7.9 | 7.9 |
| $\Sigma_0$ ($M_\odot pc^{-2}$) | 50 | 50 | 50 | 50 | 50 | 100 | 80 | 80 |
| $R_d$ (kpc) | 3.5 | 3.5 | 3.5 | 3.5 | 3.5 | 3.5 | 3.0 | 3.0 |
| $v_{tot}(R_0)$ (km/s) | 192 | 224 | 233 | 203 | 224 | 234 | 218 | 225 |
| $v_H(50)$ (km/s) | 188 | 199 | 250 | 142 | 199 | 83 | 134 | 167 |
| $v_{tot}(50)$ (km/s) | 198 | 208 | 258 | 155 | 208 | 130 | 160 | 188 |
| $\tau_{LMC}$ ($10^{-7}$) | 4.7 | 5.6 | 8.1 | 3.0 | 6.0 | 0.85 | 1.9 | 3.3 |

**Table 2:** Galactic models for LMC microlensing. Lines 2 - 8 show the model parameters: the asymptotic slope of the rotation curve ($\beta = 0$ flat, $\beta < 0$ rising, $\beta > 0$ falling), the halo flattening ($q = 1$ is spherical), the normalization velocity $v_a$, the halo core radius $R_c$, the solar distance from the galactic center $R_0$, the disk local column density ($\Sigma_0 = 50$ canonical disk, $\Sigma_0 = 100$ extreme maximal disk), and the exponential disk scale length $R_d$. Lines 9 - 12 show useful derived quantities: the total local rotation speed $v_{tot}(R_0) \approx 220$ km/sec, the rotation speed due to only the halo at 50 kpc $v_H(50\,\text{kpc})$, the total rotation speed at 50 kpc $v_{tot}(50)$, and the predicted microlensing optical depth to the LMC $\tau_{LMC}$.

To test the robustness of our results we consider the wide range of model halos shown in Table 2. Model A is the power-law-model equivalent of the simple "S" model discussed in §5.1. Model B has a very massive halo with a rising rotation curve. Model C has a relatively light halo with a falling rotation curve, while model D is similar to model A but with a halo flattened to about E6. Models A-D all have canonical disks, while models E, F and G have more massive disks and therefore substantially lighter halos. Model E is an extreme maximal disk model with a very light halo, while models F and G are "large disk" models with more realistic halo masses.

To illustrate the wide variety of models being considered, Fig. 14 shows the rotation curves (disk, halo, and total) from these models, and Fig. 11 shows the differential event rates $d\Gamma/d\hat{t}$. Using the "best" efficiencies from §4, the number of expected events $\widetilde{N}_{exp}(m)$ for a delta-function mass distribution is found using eq. 11, and the results are shown for all models in Fig. 12. The line drawn at $N_{exp} = 7.7$ marks the 95% C.L. upper limit, with points above this line being ruled out;

as discussed in §5.1, any mass distribution contained entirely within a ruled out range of masses is also ruled out.

It may be that only a fraction $f$ of the dark halo consists of Machos, the remaining $1 - f$ presumably consisting of baryonic objects outside the relevant mass range or exotic objects such as Wimps, axions, neutrinos, etc. Thus another way of displaying the results from Fig. 12 is to calculate a 95% C.L. on the Macho halo fraction, $f_{\rm lim} = 7.7/N_{\rm exp}$. Fig. 13 shows curves for $f_{\rm lim}$ for our set of models. Points above the curves are ruled out at the 95% C.L., so drawing a line at $f_{\rm lim} = 1$ will give the same excluded mass range as above, while drawing a line at $f_{\rm lim} = 0.5$ will give the mass range for which Machos contribute no more than 50% of the halo mass.

Figure 13 shows that the limits on Macho halo fraction vary substantially from model to model. For model E, the extreme maximal disk/minimal halo model, almost no useful limits $f < 1$ can be placed for any Macho mass. This model has an asymptotic rotation speed of only 83 km/sec, and is probably inconsistent with other estimates of the mass of the Milky Way halo (Zaritsky *et al.* 1989; Lin, Jones, & Klemola 1994; Freeman 1995); but we are deliberately considering an extreme range of models to test the model dependence of our results. Thus it is important to note that we find strong limits on models with massive halos (e.g. model B), and weak limits on models with light halos (e.g. model C and the large disk models). As a secondary effect, models with less massive halos have $N_{\rm exp}$ peaking at lower masses; this occurs because these halos have slower-moving Machos and produce longer event timescales for a given lens mass. Note if the halo is rotating then the relationship between Macho mass and the number of expected events will be changed, effectively shifting all the curves to the left or right. It is not known whether or not the halo is rotating, but fortunately the size of this effect is not larger than the effect of the different models.

Thus, we see that the model dependence of our limits on the Macho halo fraction is mainly due to uncertainties in the total halo mass. Therefore, we can hope to derive more model independent results by directly limiting the halo **mass** in Machos, rather than the halo **fraction**.

A useful parametrization of the mass of the halo is $v_H(50)$, the circular velocity due to the halo at a canonical distance of 50 kpc from the Galactic center. For a spherical halo, the halo mass interior to 50 kpc is simply $M_H(50) = 5.56 \times 10^{11} \, M_\odot (v_H(50)/220 \, {\rm km/s})^2$, and for non-spherical halos $M_H(50)$ can be found by direct integration of eq. 13. The quantity of interest, however, is not the halo mass, but the mass in Machos, so we define $M_{\rm lim} = f_{\rm lim} M_H(50)$ and $v_{\rm lim} = v_H(50) f_{\rm lim}^{1/2}$. These variables are convenient since the outer rotation curve of the galaxy is potentially measurable using proper motions, and is thought to be dominated by the dark halo.

Our limits $M_{\rm lim}$ for all the models are shown in Figure 15, and corresponding limits $v_{\rm lim}$ are shown in Fig. 16. It is clear that these limits are much more model-independent than the limits on $f$, even for the large disk and maximal disk models. The curves represent 95% C.L. upper limits on the amount of mass or rotation velocity Machos can contribute to the dark halo, with points above the curves being ruled out. Note that for all the models considered, objects in the $2 \times 10^{-4} - 2 \times 10^{-2} \, M_\odot$ range can contribute no more than 160 km/sec to the Milky Way rotation curve at 50 kpc (and no more than $3 \times 10^{11} \, M_\odot$ interior to this); and objects in the range $7 \times 10^{-5} \leq m \leq 0.05 \, M_\odot$ cannot contribute the entire I.A.U. value of 220 km/sec. Because of the wide range of models considered, these limits are quite robust and model independent, and they are valid whether or not our three events are due to lensing of halo objects. These are the strongest current limits on compact objects in the mass range $10^{-4} - 0.1\,\mathrm{M}_\odot$.

## 5.3: Estimate of Macho Masses and Halo Fraction

The limits of §5.2 are valid whether or not our three events are due to lensing by halo objects. However, if we make the additional assumption that our events are due to microlensing by objects in the galactic halo, then we can go beyond limits and estimate the Macho contribution to the dark halo. A sensitive way to do this is the method of maximum likelihood. A model predicts the total number of events expected and the distribution of event durations. The likelihood of a given model producing a set of $N_{\mathrm{obs}}$ detected events with durations $\widehat{t}_i$, $i = 1, \cdots, N_{\mathrm{obs}}$ is the product of the Poisson probability of finding $N_{\mathrm{obs}}$ events when expecting $N_{\mathrm{exp}}$ events, and the probability of finding the durations $\widehat{t}_i$ from the duration distribution. This can be written

$$L(m, f) = \exp(-f N_{\mathrm{exp}}(m)) \Pi_{i=1}^{N_{\mathrm{obs}}} \mu_i, \tag{14}$$

where $N_{\mathrm{exp}}$ was given in eq. 11, and $\mu_i = f E \mathcal{E}(\widehat{t}_i)(d\Gamma(\widehat{t}_i)/d\widehat{t})$. Note we are considering a delta function mass distribution here; below we consider a power-law mass function. As discussed in §4, our observed event durations are shorter than the underlying event duration when the source star is blended. In the maximum likelihood analysis, therefore we correct our observed event durations using our blending efficiency analysis. The durations used are $\widehat{t} = 38.8$ days, 21.9 days, and 31.2 days rather than the values displayed in Table 1.

For a given halo model, a smaller Macho mass gives rise to more events of shorter duration. Also a larger halo fraction will give more events, so we expect the error in determining $f$ to be correlated with the error in $m$. This is seen in Figure 17, which shows contours of log likelihood in the $m, f$ plane. The probability contours were calculated using a Bayesean method with a prior $df\,dm/m$. We use this as our standard prior hereafter, since the range of plausible values for $m$ is very large, $10^{-7}\,\mathrm{M}_\odot < m < 10^3\,\mathrm{M}_\odot$, and the $dm/m$ prior gives equal probability per decade of mass; while for $f$, the range of plausible values is $0 \leq f \leq 1$, with values near 0 or 1 being roughly equally plausible. Figure 17a is for the "standard" (model "S") halo and gives a most probable mass of $m_{2D} = 0.039\,\mathrm{M}_\odot$ and halo fraction $f_{2D} = 0.171$. (For a prior of $df\,dm$, results are shown in Fig. 17b, with best fit values of $m'_{2D} = 0.06$, and $f'_{2D} = 0.195$.) In order to estimate the Macho contribution to the halo we need the one-dimensional likelihood estimate $f_{ML}$, which we find by integrating the likelihood function over $m$. For model "S" we find $f_{ML} = 0.188^{+0.16}_{-0.10}$ for the $df\,dm/m$ prior, and $f'_{ML} = 0.216^{+0.19}_{-0.11}$ for the $df\,dm$ prior, where the errors are 68% confidence intervals. Note that the maximum likelihood results are consistent with the results given in the last section.

It is worth clarifying a difference between this analysis and that of §5.1; it appears from Fig. 17a that Macho masses of $\lesssim 10^{-4}\,\mathrm{M}_\odot$ or $\gtrsim 1\,\mathrm{M}_\odot$ are clearly excluded, while from Fig. 12 this is not the case; a model with $\sim 20\%$ of the halo in $0.05\,\mathrm{M}_\odot$ brown dwarves and the other 80% in $10\,\mathrm{M}_\odot$ black holes would be acceptable since we would not expect even 1 event from the black holes. The reason for this is that the likelihood analysis uses the extra information contained in the event timescales; however, this requires a specific model for the Macho mass function and thus a set of likelihood contours are **only** applicable to the specific mass functions under consideration, here delta functions of unique mass. Thus, the two analyses are complementary in that the event rate analysis gives rather model-independent limits while the likelihood analysis provides extra information about probable masses.

It is unlikely that the halo consists entirely of objects of the same mass, so one would like to try other mass distributions. For example, one could consider a power law mass distribution $\phi(m) = Am^\alpha$, with the normalization constant $A$ set by requiring $f = \int_{m_{min}}^{m_{max}} m \, \phi(m) \, dm$. Taking the minimum mass $m_{min}$, $\alpha$, and $f$ to be the model parameters one can create a three dimensional likelihood function. For the "S" model and the three events with durations given above, this likelihood function has a most probable slope $\alpha = -\infty$, and $m_{min}$ equal to the delta-function distribution best fit mass. That is, the likelihood function prefers the delta function mass distribution we tested above over any simple power law mass distribution; this occurs because the range in $\hat{t}$ of our 3 events is quite small, so a delta-function model is reasonable. A large sample of events will be required before we can usefully extract information on the mass function (e.g. de Rujula *et al.* 1994).

| Model | S | A | B | C | D | E | F | G |
|---|---|---|---|---|---|---|---|---|
| description | med. | med. | large | small | E6 | max disk | big disk | big disk |
| $m_{\rm ML}$ (M$_\odot$) | 0.065 | 0.050 | 0.085 | 0.031 | 0.045 | 0.007 | 0.021 | 0.032 |
| ± | +0.06 −0.03 | +0.05 −0.03 | +0.08 −0.04 | +0.03 −0.02 | +0.03 −0.02 | +0.006 −0.004 | +0.018 −0.010 | +0.03 −0.02 |
| $f$ | 0.19 | 0.16 | 0.12 | 0.31 | 0.15 | 1.1 | 0.50 | 0.29 |
| ± | +0.16 −0.10 | +0.14 −0.08 | +0.10 −0.06 | +0.26 −0.16 | +0.12 −0.08 | +0.82 −0.53 | +0.42 −0.26 | +0.24 −0.15 |
| $v_{\rm ML}(50)$ (km/s) | 82 | 80 | 86 | 79 | 77 | 88 | 95 | 90 |
| ± | +29 −25 | +29 −25 | +31 −27 | +28 −25 | +27 −22 | +28 −24 | +34 −29 | +32 −28 |
| $M_{\rm ML}(50)$ ($10^{10}$ M$_\odot$) | 7.6 | 7.4 | 8.5 | 7.2 | 6.8 | 8.9 | 10.0 | 9.2 |
| ± | +6 −4 | +6 −4 | +7 −4 | +6 −4 | +6 −3 | +6 −4 | +9 −5 | +7 −5 |
| $\tau_{\rm ML}/(10^{-8})$ | 8.8 | 9.0 | 9.6 | 9.4 | 9.0 | 9.6 | 9.6 | 9.5 |
| ± | +7 −5 | +8 −5 | +8 −5 | +8 −5 | +7 −5 | +7 −5 | +8 −5 | +8 −5 |

**Table 3:** Maximum likelihood results for the galactic models described in Table 2. The subscript ML indicates the best fit one-dimensional value and the errors are 68% C.L. found by integrating over the orthogonal variable. A Baysean method with the prior *df dm/m* was used. The variables are the best fit Macho mass $m_{\rm ML}$, the best fit halo fraction $f_{\rm ML}$, the best fit rotation speed at 50 kpc due entirely to Machos $v_{\rm ML}(50)$, the best fit "mass" in Machos interior to 50 kpc $M_{\rm ML}(50)$, and the best fit optical depth towards the LMC $\tau_{\rm ML}$.

To test the robustness of these results we next consider the delta function mass distribution for a variety of power-law halo models. Likelihood contour plots and most likely values of the mass and fraction are shown in Figure 18. Values of most probable mass and fraction are given in Table 3. As the contours show, with only three events the uncertainty in determining the lens mass is very large even within the context of simple models and a delta function mass distribution. If the Machos are in the halo, then masses in the range $0.005\,M_\odot$ and $0.2\,M_\odot$ are reasonable. Table 3 also shows the large model uncertainty in determining the halo fraction. However, the most likely halo fraction can by multiplied by the model value for the mass interior to 50 kpc, to get a more model independent estimate of the mass in Machos. Likewise the Macho contribution to the rotation velocity at 50 kpc can be found. These values are also displayed in Table 3. Remarkably we find that even the most extreme models give similar values of between 7 and 10 $\times 10^{10}\,M_\odot$ for the best fit mass in Machos interior to 50 kpc, and rotation velocities of 80 to 95 km/sec due to Machos at this distance. We see that the microlensing technique gives a fairly model independent estimate of the mass in Machos, and note that the main uncertainties in these results come from small number statistics and the assumption that all three events are due to lensing of objects residing in the dark halo. This model independence is also seen in the best fit microlensing optical depth displayed in Table 3. A range of $8.5 - 10 \times 10^{-8}$ (with very large error bars) is found, in good agreement with our 'direct' estimate $\tau_{est} = \frac{\pi}{4E} \sum \widehat{t}_i / \mathcal{E}(\widehat{t}_i) = 8 \times 10^{-8}$.

As discussed in the next section, the expected microlensing contribution from **known** populations such as the Milky Way stellar disk and spheroid and the LMC disk is less than one event, so it is quite likely that we have detected some new component of the galaxy. The question of whether or not we have detected material in the dark **halo** is made more difficult by our large rate towards the bulge (Alcock, *et al.* 1995b; Bennett *et al.* 1995). The excess events towards the LMC may be due to the same new component causing the excess bulge microlensing; this component may not be in the halo and may not, therefore, be the dark matter responsible for flat rotation curves. It is interesting to note, however, that in many scenarios put forward to explain the high microlensing rate towards the bulge, either very few events would be predicted towards the LMC (e.g. a bar or bulge-bulge microlensing), or some halo component is included (e.g. a flattened halo). These microlensing events may be the detection of a portion of the long sought dark matter.

### 5.4 Microlensing by Non-Halo Populations

It is important to note that detection of microlensing events does not require that the lens should be entirely dark, only that it should not be much brighter than the source star; thus, low-mass stars may give rise to microlensing events. Since the LMC is located at galactic latitude $b \approx -33^\circ$, the line of sight from Earth to the LMC passes mostly through the outer Galaxy, where the density of dark matter is much higher than that of known stars: thus, the event rate from known stars should constitute only a small fraction of the rate from a Macho-dominated halo. However, the ends of the line of sight pass through the disk of our Galaxy and the LMC respectively, so the predicted event rate from known stars is not negligible.

We have computed optical depths for each known component of the Galaxy, and have used a Monte-Carlo simulation to compute the distribution of event timescales, including a Scalo (1986) present-day mass function (PDMF), both neglecting and including the effects of disk rotation, and the motion of the Sun and the LMC as in Griest (1991). (Neglecting these motions lengthens the mean timescales by $\sim 20\%$, and lowers the expected event rates).

For the Milky Way thin disk, we adopt a double-exponential profile with scale height $h = 300$ pc, scale length $R_d = 4\,\text{kpc}$, local column density $\Sigma_0 = 50\,\text{M}_\odot\text{pc}^{-2}$, and 1-D velocity dispersion $\sigma = 31\,\text{km/s}$. For the thick disk, we take $h = 1\,\text{kpc}$, $R_d = 4\,\text{kpc}$, $\Sigma_0 = 4\,\text{M}_\odot\text{pc}^{-2}$, and $\sigma = 49\,\text{km/s}$. For other parameters, the optical depths scale approximately $\propto \Sigma_0 h$, and the mean timescales scale approximately as $\sqrt{h}/\sigma$; these values are almost independent of $R_d$ since most of the disk lensing occurs near $r \approx R_0$.

For the spheroid, we adopt a density profile $\rho(r) = 1.18 \times 10^{-4}(r/R_0)^{-3.5}\,\text{M}_\odot\text{pc}^{-3}$, and a 1-D velocity dispersion of $120\,\text{km/s}$. (This density profile clearly must be cut off at small $r$, but this is irrelevant here since the LMC sight-line is always at $r > 0.99 R_0$.)

For the LMC disk, we adopt a double exponential profile with $h = 250$ pc, a central surface brightness of $140 L_\odot \text{pc}^{-2}$, $M/L = 3$ and inclination $i = 30^\circ$, which gives a central face-on column density of $363\,\text{M}_\odot\text{pc}^{-2}$. We take a 1-D velocity dispersion of $25\,\text{km/s}$ for both sources and lenses, and we average over the depth of the LMC assuming that the source stars are distributed with the same profile, and no extinction through the LMC. We compute the LMC observables both at the center, and using an average (weighted by stellar density) over the locations of our fields; using a scale length of $R_d = 1.6\,\text{kpc}$, the averaged optical depth is 60% of that at the center.

For each component we compute the optical depth $\tau$, the distribution of event timescales for a Scalo PDMF, and the expected number of events $N_{\text{exp}}$ using our standard efficiency, shown in Table 4. For comparison, we show the values for a dark halo with the same Scalo PDMF used for the visible components. Our estimates for $\tau$ and $\langle \hat{t} \rangle$ are comparable to those of Gould *et al.*(1994) and Giudice *et al.*(1994), accounting for the different parameter choices.

We see from the Table that the spheroid and thick disk produce almost negligible lensing rates, unless either has a dark component far above that in visible stars (Gould *et al.*1994, Guidice *et al.*1994). However, the thin disk and LMC disk produce significant lensing rates. Our thin-disk values are appropriate for the old component; the younger component has a smaller $h$ and $\sigma$, so we will somewhat overestimate the true rate; also, we ignore the fact that part of the disk column density is in gas, and part in bright stars which will not cause detectable lensing. The optical depth at the center of the LMC disk is greater than that from the Milky Way disk, due to the much higher column density which outweighs the smaller inclination angle.

Adding the contributions from the thin and thick disks, the spheroid and the averaged LMC value, we estimate that the expected number of detected events in our dataset from lensing by all known stellar populations is $N_{\text{exp}}(\text{stars}) \approx 0.55$. Thus, we conclude that if all 3 candidate events are genuine microlensing, there is a significant excess above the expectations from stellar lensing, at $\approx 98\%$ confidence. If 2 of our candidates are microlensing, the significance would be modest, $\approx 90\%$; while if only event 1 is microlensing, this could reasonably be accounted for by known stellar populations.

Also, the timescales of the observed events are somewhat shorter than expected from stellar lensing, though not dramatically so; with the above parameters, we find that only 17% of detected

| Component | $\sigma$ | $\tau$ ($10^{-7}$) | $\langle\hat{t}\rangle_A$ (days) | $\langle\hat{t}\rangle_B$ (days) | $\langle l\rangle$ (kpc) | $\Gamma$ ($10^{-7}$yr$^{-1}$) | $N_{\rm exp}$ |
|---|---|---|---|---|---|---|---|
| Thin disk | 31 | 0.15 | 144 | 112 | 0.96 | 0.62 | 0.13 |
| Thick disk | 49 | 0.036 | 157 | 105 | 3.0 | 0.16 | 0.034 |
| Spheroid | 120 | 0.029 | 105 | 95 | 8.2 | 0.14 | 0.030 |
| LMC center | 25 | 0.53 | 93 | 93 | 49.8 | 2.66 | (0.58) |
| LMC avge | 25 | 0.32 | 93 | 93 | 49.8 | 1.60 | 0.35 |
| Halo S | 155 | 4.7 | 95 | 89 | 14.4 | 24.3 | 5.5 |

**Table 4:** Microlensing quantities for various galactic components described in the text: the adopted 1-D velocity dispersion $\sigma$, optical depth $\tau$, and average event timescale $\langle\hat{t}\rangle$ for lenses with Scalo PDMF. (For a delta-function PDMF at $1\,{\rm M}_\odot$, timescales are longer by a factor 1.5 ). The two values $\langle\hat{t}\rangle_A$ and $\langle\hat{t}\rangle_B$ respectively ignore and include disk rotation and Sun and LMC motion; the latter are more realistic, but we include the former for comparison. $\langle l\rangle$ is the mean lens distance, including motions. The theoretical event rate $\Gamma = 4\tau/\pi\langle\hat{t}\rangle_B$. The expected number of events $N_{\rm exp}$ includes our exposure $E = 9.7 \times 10^6$ star-yr and our detection efficiency averaged over the $\hat{t}$ distribution. For the LMC two rows are shown, firstly at the center and secondly averaged over our fields using a 1.6 kpc scale length; the $N_{\rm exp}$ value for the center is not applicable, so is shown in brackets.

LMC events and 8% of disk events would be shorter than $\hat{t} < 35$ days, and only 4% and 1.5% of LMC and disk events would be shorter than $\hat{t} < 20$ days.

This conclusion may seem surprising since it was estimated by Sahu (1994) that LMC stars could almost entirely account for the observed optical depth, with a value $\tau \approx 0.5 \times 10^{-7}$ across the LMC bar. This was disputed by Gould (1994), who finds a relation between the optical depth of a self-gravitating disk and its line-of-sight velocity dispersion of

$$\tau \leq 2\langle v^2\rangle/c^2 \sec^2 i \qquad (15)$$

For the observed velocity dispersion $\langle v^2\rangle^{1/2} \approx 25\,{\rm km/s}$, $\tau \lesssim 0.1 \times 10^{-7}$. This may suggest that the scale-height of the LMC is smaller than estimated here. Part of the difference appears to arise from the fact that Sahu assumes a constant mass to light ratio for the whole LMC, including a probable dark matter contribution; this gives a rather extreme surface density of $1100\,{\rm M}_\odot{\rm pc}^{-2}$ for the LMC bar. Another estimate is given by Wu (1994) who finds $\tau \approx 0.33 \times 10^{-7}$ at the center of the LMC disk using more conventional parameters.

However, the main difference in our conclusions arises from our use of the **event rate** rather than the optical depth. As is shown by Han & Gould (1994), the uncertainty in the optical depth is considerably larger than given by Poisson statistics due to the wide range of event timescales. Thus a simple comparison of optical depths may well be misleading: the probability of observing a **single** event with $\hat{t} \gtrsim 80$ days from lensing by a star is $\sim 0.5$, and such an event would in fact give a similar optical depth estimate to that given here. However, the probability of observing $N_{\rm obs} \geq 3$ events from lensing by stars appears to be low, $\sim 2\%$, and the short event timescales also disfavour this interpretation.

Another feature of Table 4 is that the various components produce quite similar mean timescales; this has been previously noted by Gould *et al.*(1994), and is due to the fact that the more extended components necessarily have larger velocity dispersions. This is useful in that the mass estimates are not very sensitive to the assumed lens population, but means that it will be very hard to pin down the lensing component based on timescales alone.

## 6. Discussion

The results presented here provide some very interesting constraints on the nature of the dark halo of our Galaxy; though they are limited somewhat by small-number statistics and uncertainties in the total mass of the halo. The observed number of 3 microlensing candidates is rather unexpected, and makes our results quite hard to interpret definitively; if we had observed either $N \leq 1$ or $N > 10$ candidate events, our conclusions would have been much clearer.

Our most definite conclusion is that objects of around Jupiter to brown dwarf mass, from $\sim 3 \times 10^{-4}$ to $0.06 \, {\rm M}_\odot$, cannot comprise the dominant component of the **standard** halo (eq. 4): they comprise $< 50\%$ of such a halo at 95% confidence level. For a standard halo, we are also able to limit the contribution of brown dwarfs of $M = 0.1 \, {\rm M}_\odot$; these comprise $< 66\%$ of the standard halo at 95% confidence.

For the specific halo model considered, these limits are quite robust; although there are some systematic uncertainties in our efficiency estimates, our Monte-Carlo simulations provide a full end-to-end test of both the photometric and analysis software, and no unexplained loopholes are found. [ Furthermore, we have now analysed a large sample of bulge data, and the event rate is unexpectedly high. This would be very hard to explain in plausible mass models of the disk & bulge if our efficiency had been seriously overestimated.] We have explored a range of assumptions for estimating our detection efficiency, and find that this can account for at worst a $\sim 20\%$ error from our 'best' efficiency measurement.

However, as is shown by Alcock *et al.* (1995c), there are substantial uncertainties in the density profile and total mass of the dark halo, which cause a corresponding uncertainty in the predicted microlensing rates. We have explored a large set of realistic halo models based on those of Evans (1993), which include the mass of the galactic disk and span the entire range of observationally allowed values for the disk and halo mass; we find a tight correlation between the mass in Machos interior to 50 kpc and the predicted number of microlensing events.

Thus, if the real halo is at the lower-mass end of the observationally allowed range, it could be comprised mostly of brown dwarfs in the $0.03 - 0.08 \, {\rm M}_\odot$ range and still be entirely consistent with our results. The fact that our 3 microlensing candidates are difficult to explain as lensing by known stars makes this an interesting possibility, but by no means a unique one.

At present, there appear to be at least 5 plausible hypotheses to account for our results, which provide several testable predictions as follows:

H1) All three observed events are microlensing by halo brown dwarfs, but these contribute only $\sim 25\%$ of the total halo mass. The rest could be contributed by more massive objects e.g.

white dwarfs or black holes, or by elementary particle dark matter such as WIMPs, axions, or massive neutrinos. The presence of a significant amount of baryonic dark matter is expected in models dominated by massive neutrinos or cold dark matter (CDM). Neutrino dominated models require baryonic dark matter to dominate dwarf galaxy halos while in CDM-dominated models, between $10 - 50\%$ of the inner regions of dark halos should be baryonic, depending on the amount of dissipation (e.g. Gates, Gyuk & Turner 1995). In this scenario, the outer rotation curve should be close to flat, and the future microlensing rate should be similar to that found here.

H2) All three observed events are microlensing, and the halo is dominated by brown dwarfs, but has a mass interior to the LMC of only $\sim 2 \times 10^{11} \, M_\odot$. This would predict that the rotation curve of the outer Galaxy should be falling to $\sim 160 \, \text{km/s}$ at $50 \, \text{kpc}$; and that the microlensing rate in future data should be somewhat higher than found here.

H3) All three observed events are microlensing, but arising from substellar objects in a population other than the dark halo. A thin disk probably cannot produce sufficient events without exceeding the mass limits from the rotation curve, but dark objects in a thick disk (Gould *et al.* 1994), or a spheroid (Guidice *et al.* 1994) would be consistent. This would be quite hard to discriminate from H1 without parallax measurements.

H4) Two or three observed events are microlensing, but arising from a statistical fluctuation of the lensing rate by known stars. This explanation is statistically unlikely, though would be formally allowed at 90% confidence with 2 microlensing events; but it is also somewhat disfavoured by the observed timescales. This would predict that very few candidates should be observed in future, with a considerably longer mean timescale.

H5) Only the best event is genuine microlensing, and events 2 & 3 represent a previously unknown class of variable star. In this case, event 1 could be reasonably explained as arising from microlensing by a star in the LMC or Milky Way disk, merely with a higher amplification and shorter timescale than average. This would predict that very few 'good' lensing candidates should occur in future data, with an excess of low-amplitude variable-star candidates; these latter may be expected to show spectroscopic peculiarities or deviations from the theoretical shape, which should be detectable using our real-time alert system.

Note that if hypotheses H3, H4 or H5 were correct, this would suggest that our limits on Machos in §5.2 are too conservative, and the true limits would be stronger by approximately a factor $3/7.7$; thus Machos from $3 \times 10^{-4}$ to $0.1 \, M_\odot$ could contribute no more than $8 \times 10^{10} \, M_\odot$ to the halo mass.

The prospects are very good for improving the conclusions presented in this paper. The dataset analysed here represents only $\approx 1/3$ of the LMC frames taken to date, and our strategy during the third year was slightly modified to increase the number of stars monitored for events with $\hat{t} > 20$ days. As the timespan of observations increases, we will become sensitive to events with timescales $\gtrsim 100$ days, constraining the contribution of black holes and stellar remnants to the halo. We are also accumulating a useful quantity of data on the SMC; the ratio of LMC/SMC event rates can provide useful constraints on the distribution of microlensing objects (Sackett & Gould 1994, Alcock *et al.* 1995c).

We have now implemented a prototype "real-time" event detection system, which is currently running on a subset of our fields. This system has produced two events in 1994, and eight more in April-May 1995. The events discovered in 1994 comprised one towards the bulge on 1994 Aug 31 and one in the LMC on 1994 Oct 14 (Alcock *et al.*, IAU Circulars 6068, 6095). Both were announced around a week prior to peak brightness; thus frequent images were taken at several telescopes, and both were observed spectroscopically near peak amplification. The lightcurves show a good fit to microlensing, and the spectra show no peculiarities. (Benetti *et al.* 1995; E. Giraud, IAUC 6097; C. Joseph, private communication; unpublished MACHO collaboration spectrum). For the LMC event, the timescale was $\hat{t} = 42$ days, and the peak amplification was $A_{\max} = 2.98$. Although this event does not fall within the timespan used here, thus cannot be added to the statistics, it does tend to support the microlensing interpretation for our 3 earlier events. If it proves possible to detect most future events in real-time, follow-up observations will enable the data quality to be greatly improved relative to events 2 & 3, which should give a reasonably unambiguous test of microlensing.

The definitive method for determining the population of the lenses is to carry out follow-up observations from a distant satellite (Gould 1994a,b), which measures the velocity of the lens projected to the solar system and can cleanly discriminate between disk, halo and LMC lenses on an event-by-event basis (as well as providing a conclusive proof of microlensing). The real-time detection capability is at least a first step in this direction.

We are currently investigating a distinct analysis code designed to search the current dataset for very short events with $\hat{t} < 2$ days; while this could not produce a convincing detection of events with the current sampling, it appears to be capable of setting interesting upper limits on the abundance of very low-mass Machos with $10^{-6} < m < 10^{-4} \, M_\odot$. An upper limit in a similar mass range from the EROS CCD experiment has recently been presented by Aubourg *et al.*(1995). Thus, if short timescale events continue to be absent, it appears that the entire mass range from $10^{-7}$ to $\sim 0.02 \, M_\odot$ may soon be excluded as the dominant component of the halo. This is of major importance since such objects would not be observable by any other current technology. If the halo objects are just below $0.08 \, M_\odot$ as hinted by current results, those in the solar neighbourhood may be accessible to the ISO survey, and should be easily detectable by SIRTF.

Although the number of detected events is still small, the 'model' uncertainties in the predicted event rate provide a major source of uncertainty in our conclusions. This situation could be greatly improved by new measurements of the dynamics of the outer Galaxy, since the predicted microlensing event rate is closely correlated with the halo mass in Machos interior to 50 kpc. Radial velocities for large samples of distant spheroid stars, and especially proper motion measurements of satellite galaxies (e.g. Lin *et al.* 1995) are extremely desirable for this reason: if the LMC is in an approximately circular orbit with velocity $v_c = 220 \, \text{km/s}$, then the dark halo cannot be dominated by substellar objects in the mass range $3 \times 10^{-4} \, M_\odot < m < 0.06 \, M_\odot$.

ACKNOWLEDGEMENTS

We are very grateful for the skilled support given our project by the technical staff at the Mt. Stromlo Observatory, in particular our dedicated observer, Simon Chan. We are very grateful to Joe Silk for suggesting the initial collaboration and to Bernard Sadoulet for early and continued support.

Work performed at LLNL is supported by the DOE under contract W7405-ENG-48. Work performed by the Center for Particle Astrophysics on the UC campuses is supported in part by the Office of Science and Technology Centers of NSF under cooperative agreement AST-8809616. Support from the Bilateral Science and Technology Program of the Australian Department of Industry, Science, and Technology is gratefully acknowledged. KG acknowledges DOE OJI, Sloan, and Cottrell Scholar awards. CWS thanks the Sloan and and Packard Foundations for their generous support. WJS acknowledges use of Starlink computing resources at Oxford.

# Figure Captions

Figure 1      A flow chart of the procedure for selection of microlensing candidates.

Figure 2      The locations of lightcurves passing cuts (1)-(5) in the $(A_{\max}, \Delta\chi^2)$ plane. The symbols are described in §3.

Figure 3      The lightcurve of a star showing a glitch due to the optical misalignment near day 98. For each star the two panels show the red and blue passbands. Flux is on a linear scale, in units of the observed median flux. Time is in days from Julian Day $2449000 \equiv 1993$ Jan 12.5 .

Figure 4      The lightcurves of typical short and long timescale 'bumpers' described in §3.1. Units as in Figure 3.

Figure 5      A 5.4 x 5.4 arcmin red image centered at (5:35:22.9, -69:15:15 [J2000]). Boxes mark the positions of the "stars" which show events resembling microlensing. Small boxes contain one of these "stars"; large boxes contain two or three (14 total).

Figure 6      The lightcurves of two typical variables from the region of high nebulosity. Units as in Figure 3.

Figure 7      The lightcurves of the three candidate microlensing events. The solid lines show the best-fit microlensing curves, with the parameters shown in Table 1. Units as in Figure 3.

Figure 8      Recovery of parameters for simulated microlensing events in randomly selected lightcurves. (a) shows contours of equal likelihood for the fractional errors in recovered parameters $u_{\min}, \widehat{t}$ using the procedure described in § 4.1. (b) shows contours for artificial events incorporating blending effects described in § 4.2.

Figure 9      MACHO Year 1 microlensing event detection efficiency $\mathcal{E}$ as a function of event duration $\widehat{t}$. The left axis is labeled with event detection efficiency per monitored object, and the right axis gives the efficiency–corrected exposure. The upper curve takes into account only the temporal sampling effects of § 4.1 while the rest correct for event degradation from blending by using different stellar luminosity functions to estimate the increased number of sources. The best estimate is the solid curve.

Figure 10      Luminosity function models used in the efficiency estimates. The lower curve is the observed LF for all fields analysed. The upper curve is an estimate of the underlying stellar LF. The intermediate curve is the measured LF of a relatively

| | |
|---|---|
| Figure 11 | Theoretical differential event rates as a function of $\hat{t}$ for models described in Table 2. A delta-function Macho mass function at $1\,M_\odot$ is assumed; for other masses one simply scales the $\hat{t}$-axis by $\sqrt{m}$ and the rate-axis by $m^{-1}$, hence the total event rate scales as $m^{-0.5}$. The thick solid line shows the "standard" model S, the thin solid lines show models B (highest),A,D,C (lowest). The short-dashed line shows the maximal disk model E, and the long-dashed lines show the large disk models F,G. |
| Figure 12 | Number of expected events as a function of the Macho mass for the models described in Table 2. The lines are coded as in Fig. 11. The horizontal line drawn at $N_{\mathrm{exp}} = 7.7$ indicates the 95% C.L. upper limit for Poisson statistics given that we have seen no more than 3 microlensing events. |
| Figure 13 | 95 % C.L. upper limits on the Macho halo fraction as a function of the Macho mass for the models described in Table 2. The lines are coded as in Fig. 11. |
| Figure 14 | Theoretical rotation curves for the models described in Table 2. The dashed line shows the disk contribution; the lower and upper solid line are the halo contribution and the total. |
| Figure 15 | 95% C.L. upper limits on the mass in Machos interior to 50 kpc as a function of the Macho mass for the models described in Table 2. The lines are coded as in Fig. 11. The horizontal line at $5.56 \times 10^{11}\,M_\odot$ shows the mass in a spherical halo with the I.A.U. rotation value of 220 km/sec at 50 kpc. |
| Figure 16 | 95% C.L. upper limits on the rotation speed at 50 kpc due to Machos as a function of the Macho mass for the models described in Table 2. The lines are coded as in Fig. 11. The horizontal line at 220 km/sec shows the I.A.U. value assuming a flat rotation curve out to 50 kpc. |
| Figure 17 | Likelihood contours for halo model S with a delta function mass distribution. The positions of the most likely Macho mass $m_{\mathrm{ML}}$ and most likely Macho halo fraction $f_{\mathrm{ML}}$ are marked with a +. Part (a) shows the contours for the prior $dfdm/m$ (logarithmic in $m$), while part (b) shows contours for the the prior $dfdm$ (uniform in $m$). |
| Figure 18 | Likelihood contours for halo models S, and A-G with a delta function mass distribution. The positions of the most likely Macho mass $m_{\mathrm{ML}}$ and most likely Macho halo fraction $f_{\mathrm{ML}}$ are marked with a +. The models are described in Table 2. |

**Appendix A: Distribution of event durations**

The distribution of event durations is determined by the density distribution $\rho(\vec{r})$, and velocity distribution $f(\vec{v})$ of Machos in the Galaxy. For a spherical halo and an isotropic velocity distribution, eq. 10 of Griest (1991) gives the differential microlensing rate

$$d\Gamma = \frac{\rho(r) u_T r_E L}{\pi m v_c^2} e^{-(\mathbf{v_r} - \mathbf{v_t})^2/v_c^2} v_r^2 \cos\theta dv_r d\theta d\alpha dx, \tag{A1}$$

where $L \approx 50$ kpc is the distance to the LMC, $x$ is the distance between the Earth and the Macho in units of $L$, $r$ is the distance from the Macho to the Galactic Center, $v_c \approx 220$ km/sec is the solar rotation velocity, $\mathbf{v_r}$ is the transverse velocity of the Macho, $\mathbf{v_t}$ is the transverse velocity of the line-of-sight (set equal to zero in this paper), and the angles are defined in Griest 1991. By changing variables, Griest (1991) transformed this differential rate into the distribution of event durations using $t_e$, the time for which the amplification is greater than the threshold amplification, but the distribution in terms of the more useful $\hat{t}$ was not given. (Note, $\hat{t} = t_e(u_T^2 - u_{\min}^2)^{-1/2}$ is the time to cross the Einstein diameter, where $u_T$ is the threshold impact parameter, and $u_{\min}$ is the minimum impact parameter, both in units of the Einstein radius $r_E$.) While the averages of the $\hat{t}$ and $t_e$ distributions are simply related, $\langle t_e \rangle = \frac{\pi}{4} \langle \hat{t} \rangle$, the distributions themselves differ. Using $\hat{t} = 2r_E u_T/v_r$ and $xL$ as the distince between us and the Macho, we find

$$\frac{d\Gamma}{d\hat{t}} = \frac{32 L}{m \hat{t}^4 v_c^2} \int_0^{x_h} \rho(x) r_E^4(x) e^{-Q(x)} e^{-(v_t(x)/v_c)^2} I_0(P(x)) dx, \tag{A2}$$

where $x_h \approx 1$ is the extent of the halo, $Q(x) = 4r_E^2(x) u_T^2/(\hat{t}^2 v_c^2)$, $P(x) = 4r_E(x) u_T v_t/(\hat{t} v_c^2)$, $r_E^2(x) = 4Gmx(1-x)L/c^2$, $I_0$ is a Bessel function, and $u_T = 1$ is the threshold impact parameter. For the purposes of this paper we set $v_t$, the transverse LMC and Earth motion to zero, simplifying this equation to that given §5.1, eq. 10.

Specializing to the model of Griest (1991) with a core radius of $R_c \approx 5$ kpc, and a radius of the solar circle $R_0 \approx 8.5$ kpc, we have

$$\frac{d\Gamma}{d\hat{t}} = \frac{512 \rho_0 (R_c^2 + R_0^2) L G^2 m}{\hat{t}^4 v_c^2 c^4} \int_0^{x_h} \frac{x^2(1-x)^2}{(A' + Bx + x^2)} e^{-Q(x)} e^{-(v_t(x)/v_c)^2} I_0(P(x)) dx, \tag{A3}$$

where $A' = (R_c^2 + R_0^2)/L^2$, $B = -2(R_0/L) \cos b \cos l$, $b$ is galactic latitude, and $l$ is galactic longitude. Note that for the parameters chosen, the rotation velocity at the solar circle due to the halo is 127 km/sec, with a canonical disk contributing 143 km/sec, for a total of 192 km/sec. While this is not quite the standard value, it is close, especially if one allows for the bulge and a somewhat larger than canonical disk. We use this model since it has been used in the past. See Evans & Jijina (1993), for more discussion of the effect of the disk on microlensing.

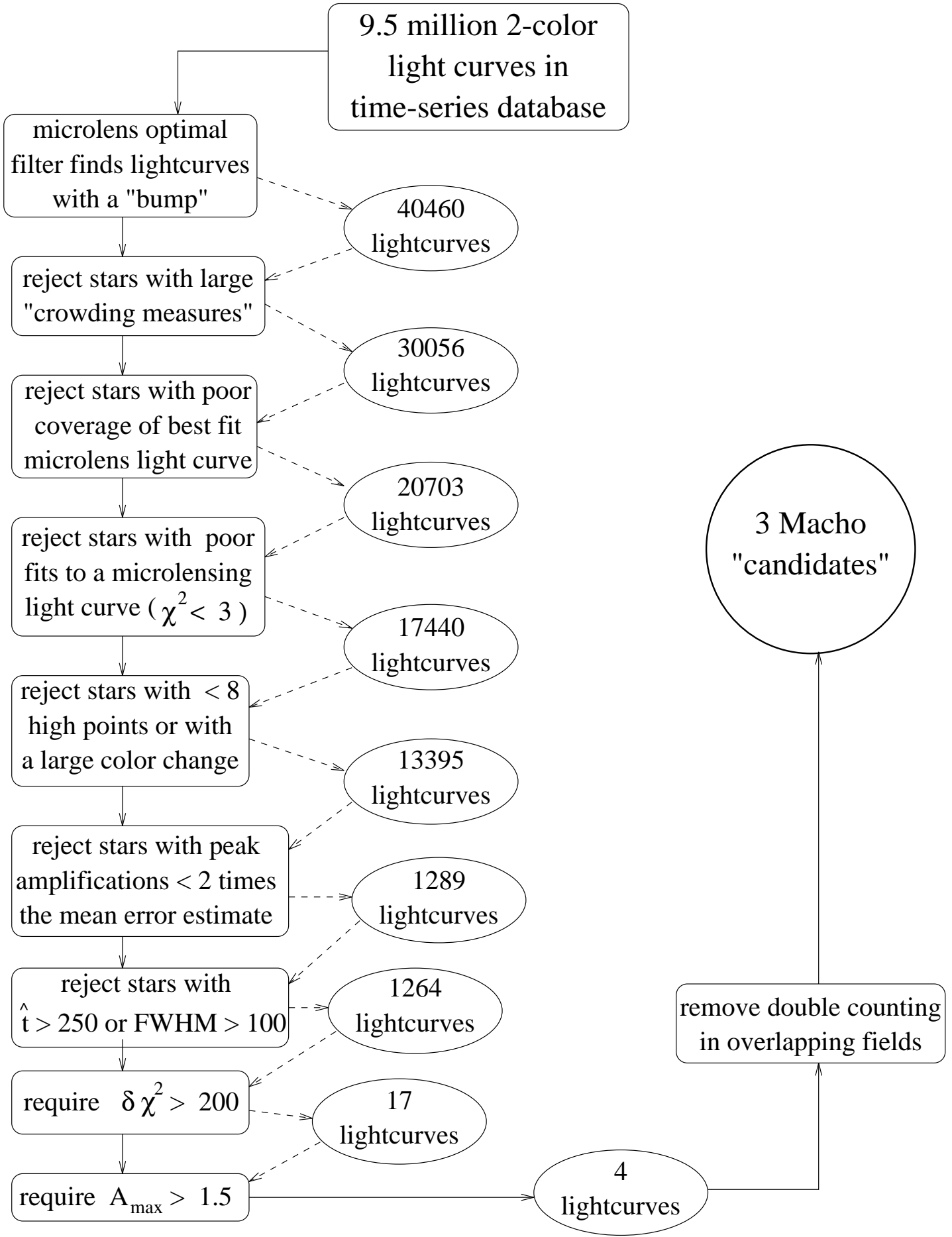

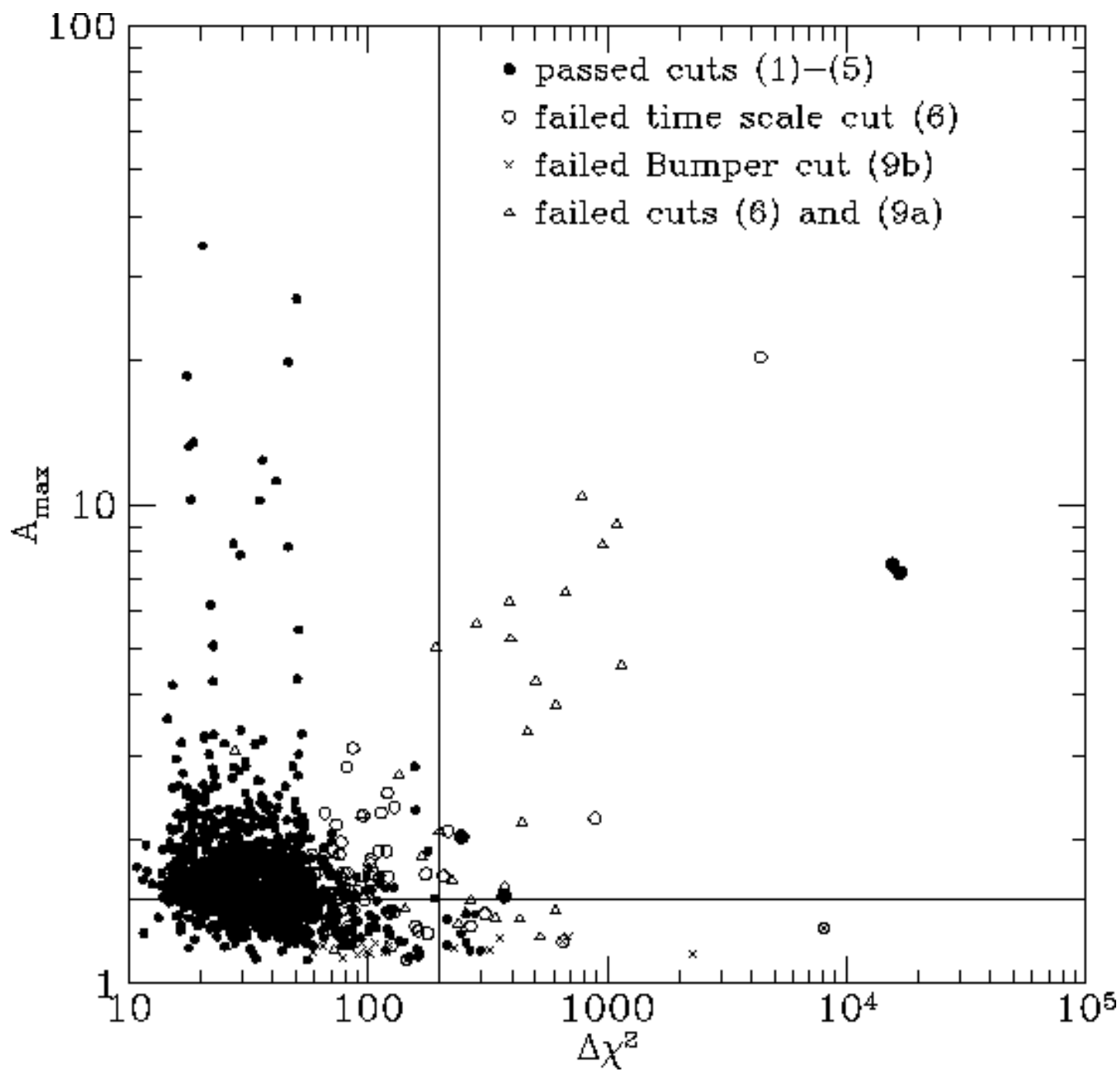

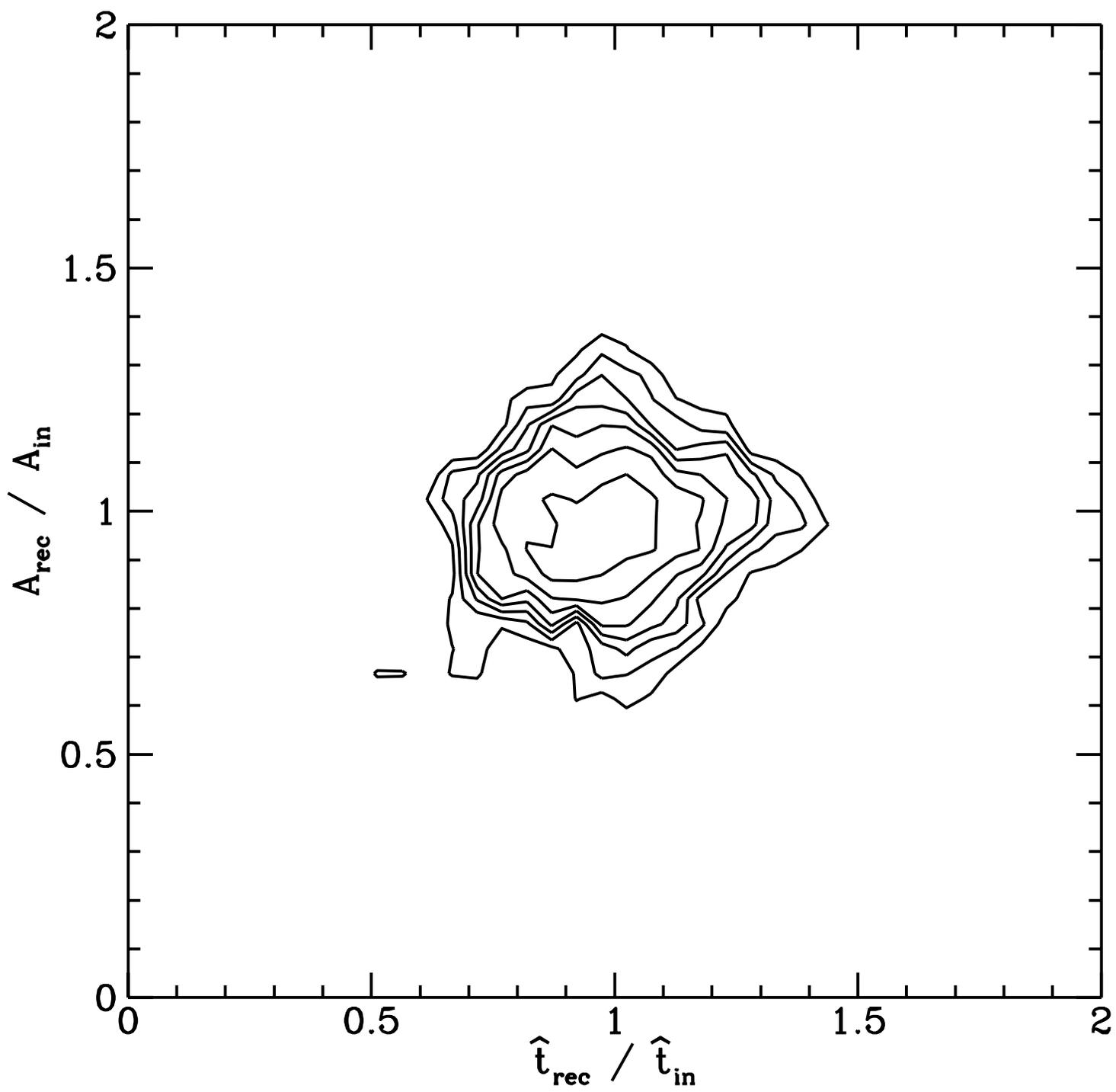

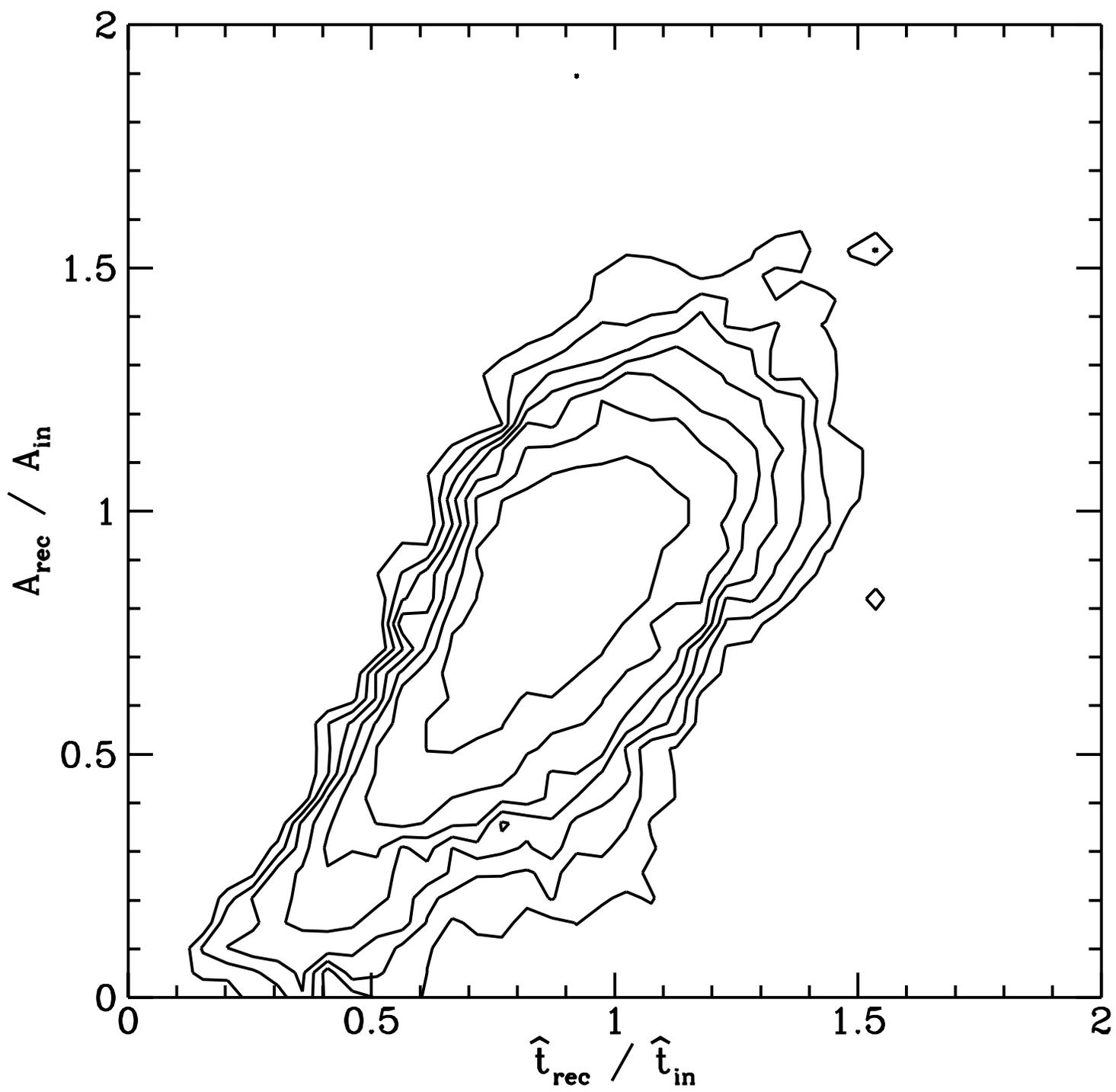

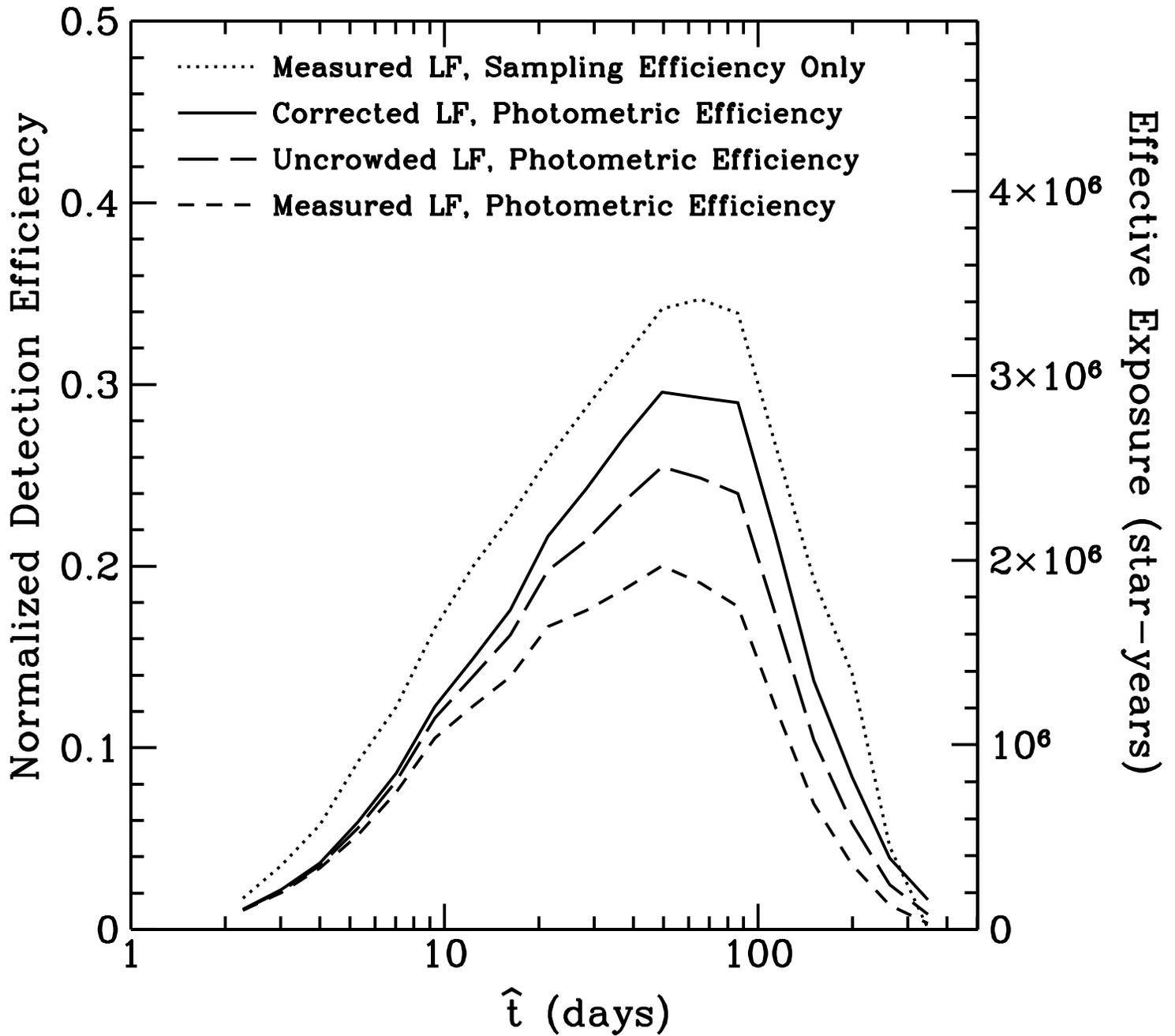

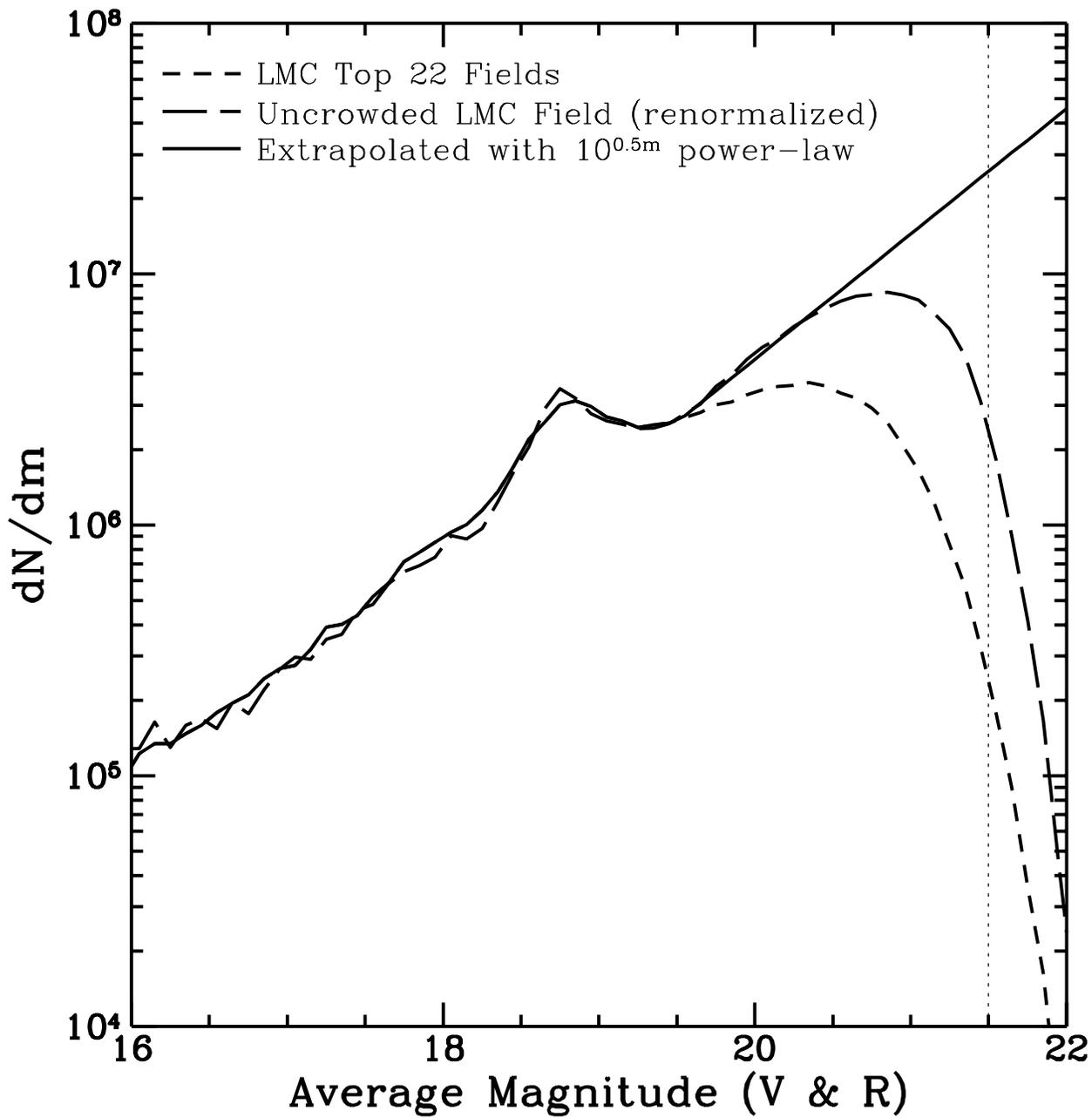

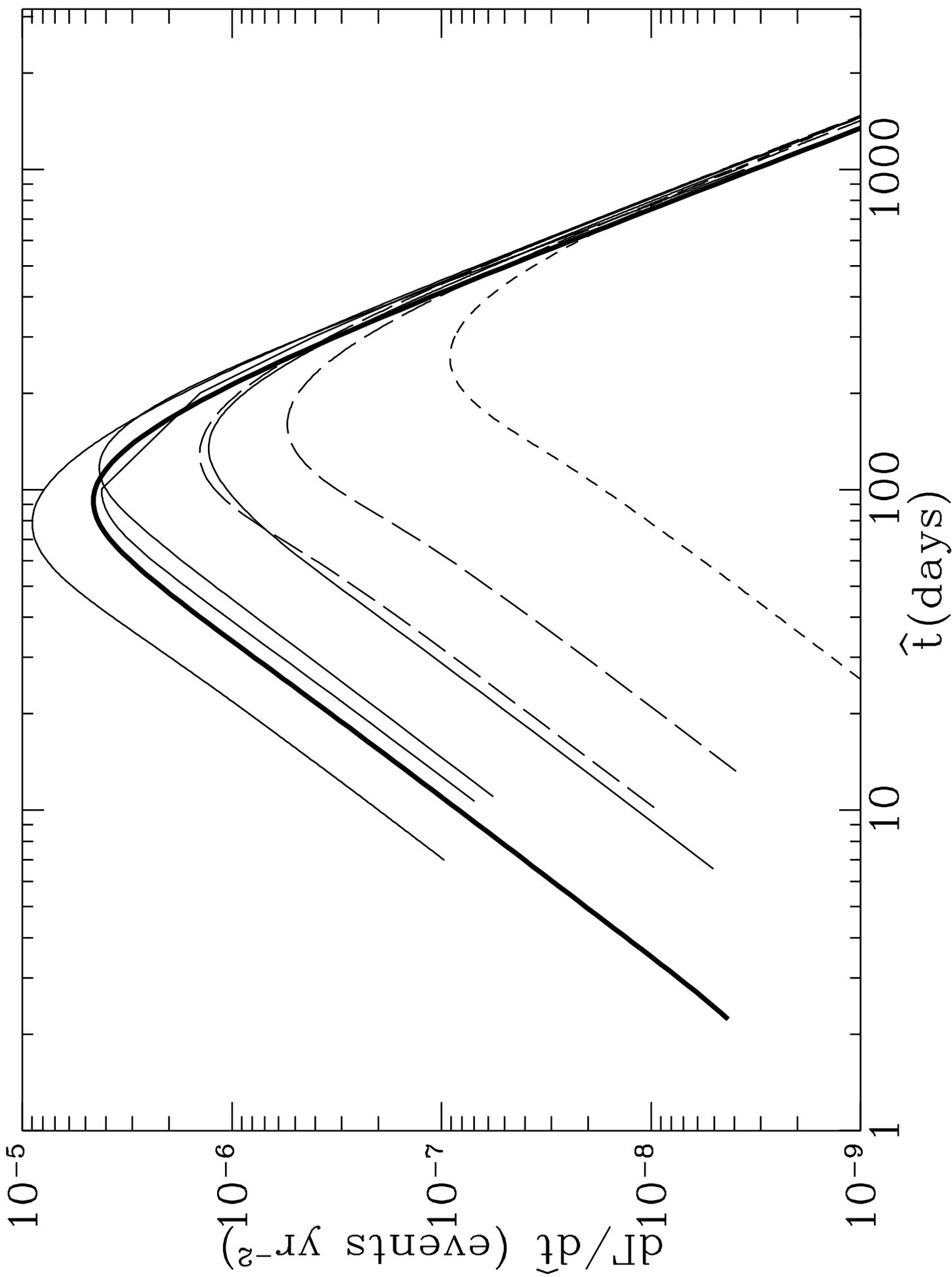

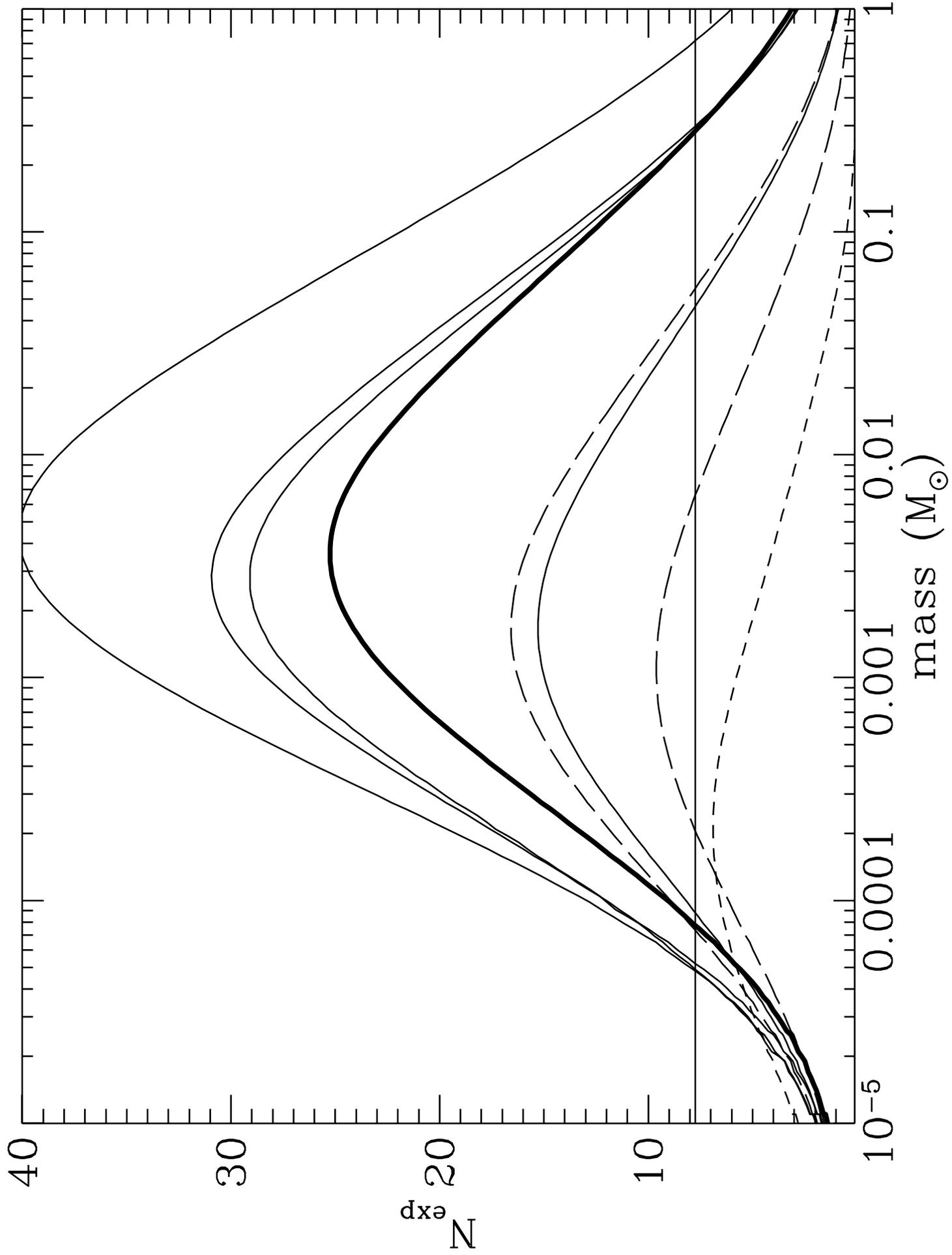

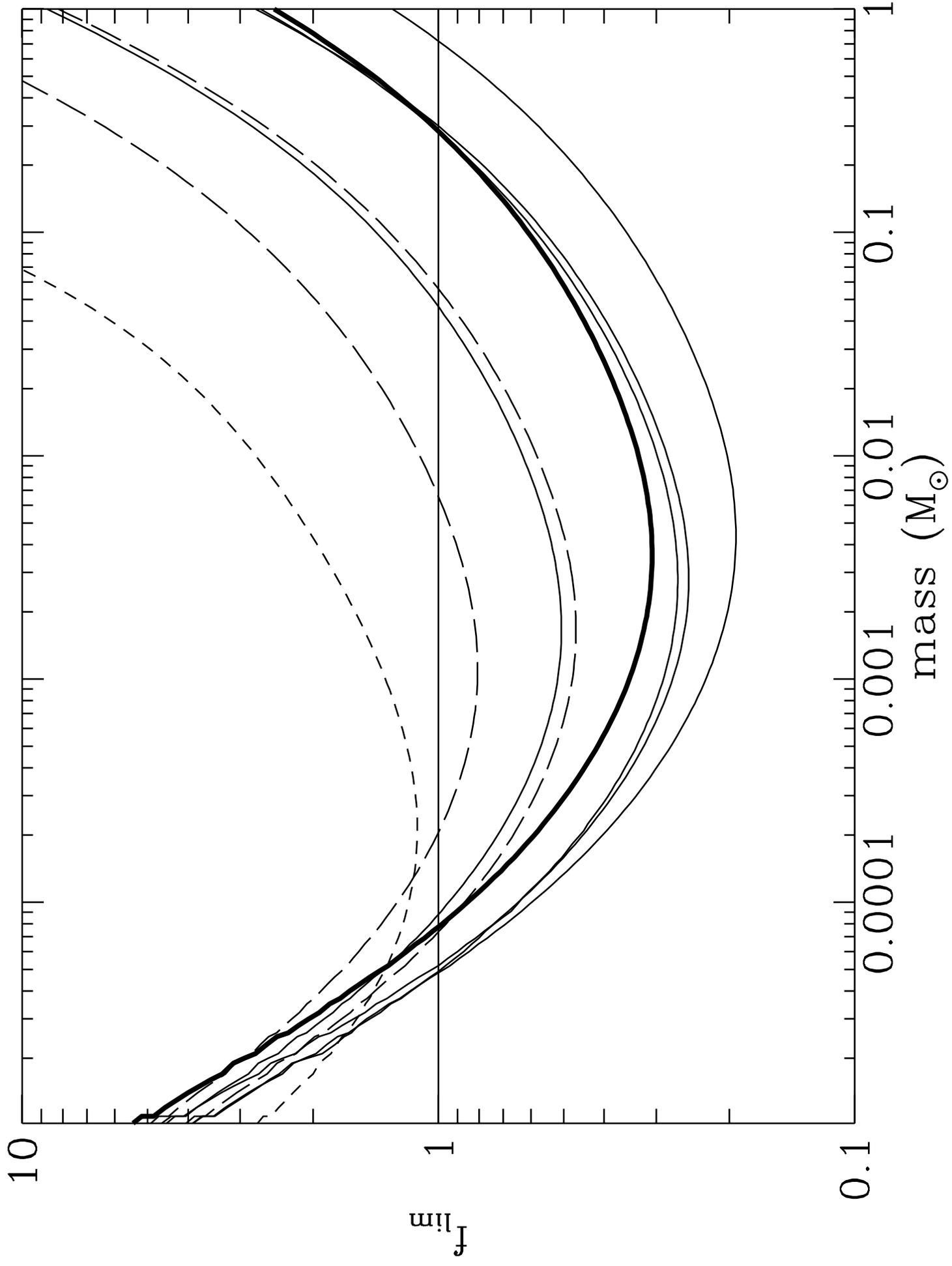

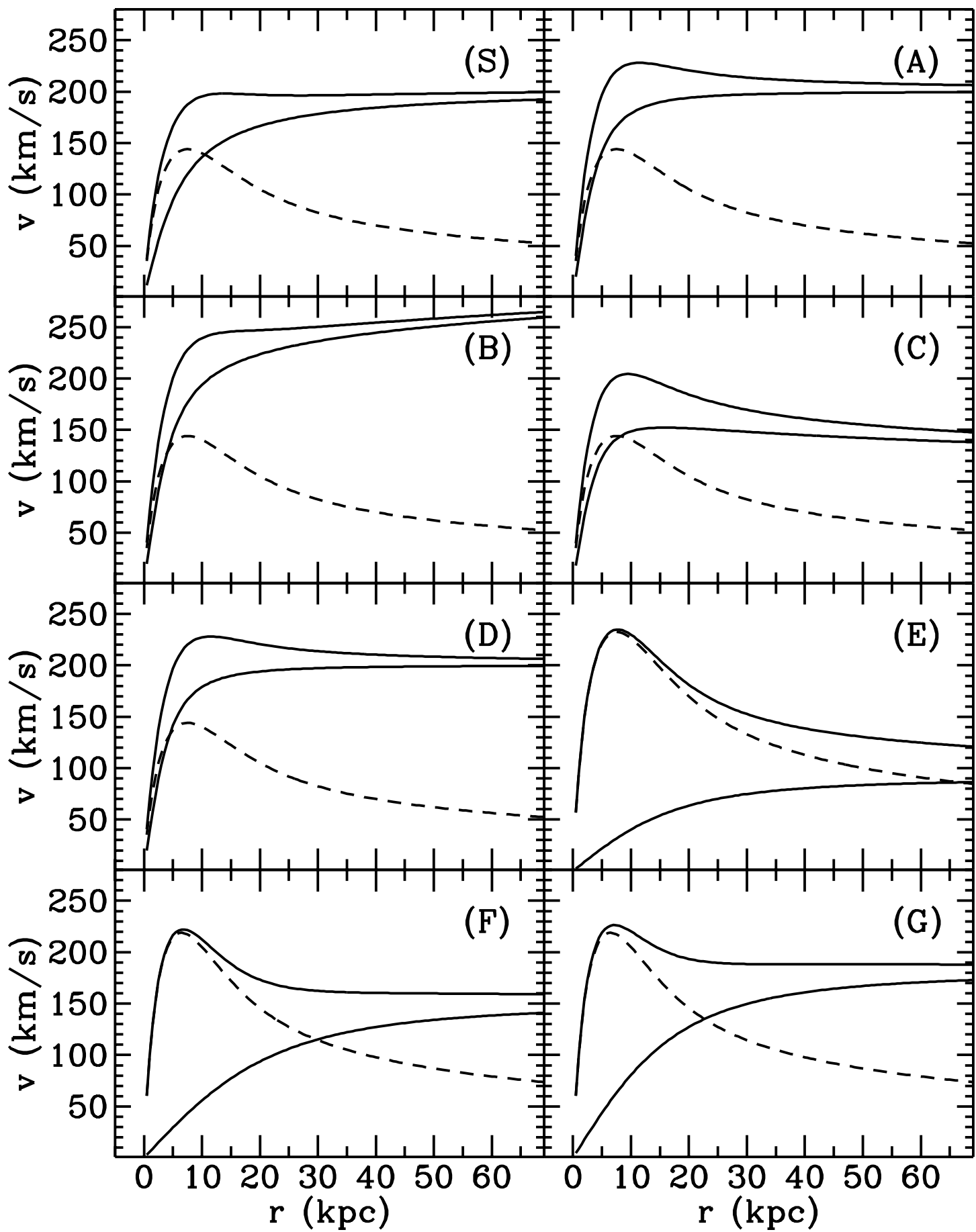

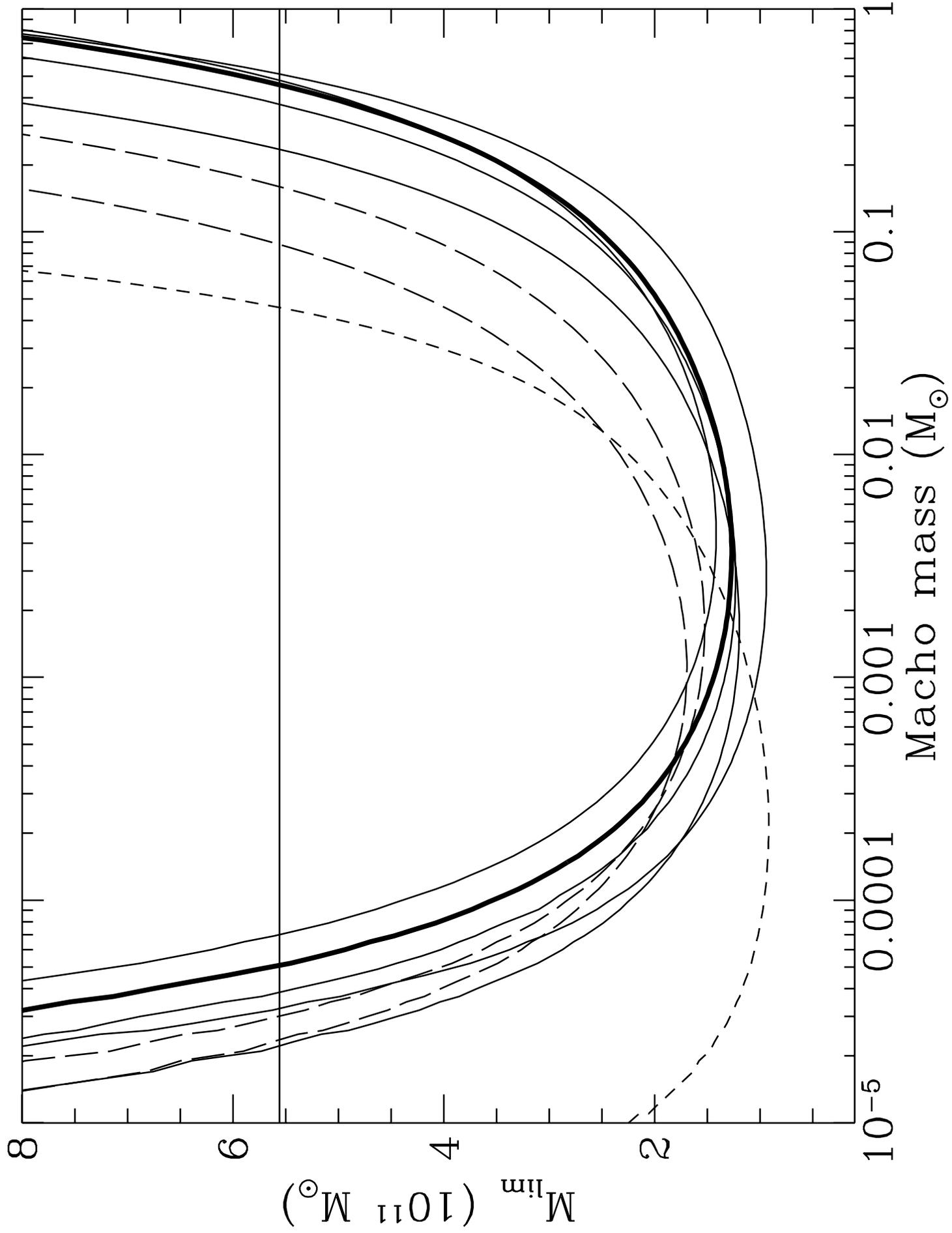

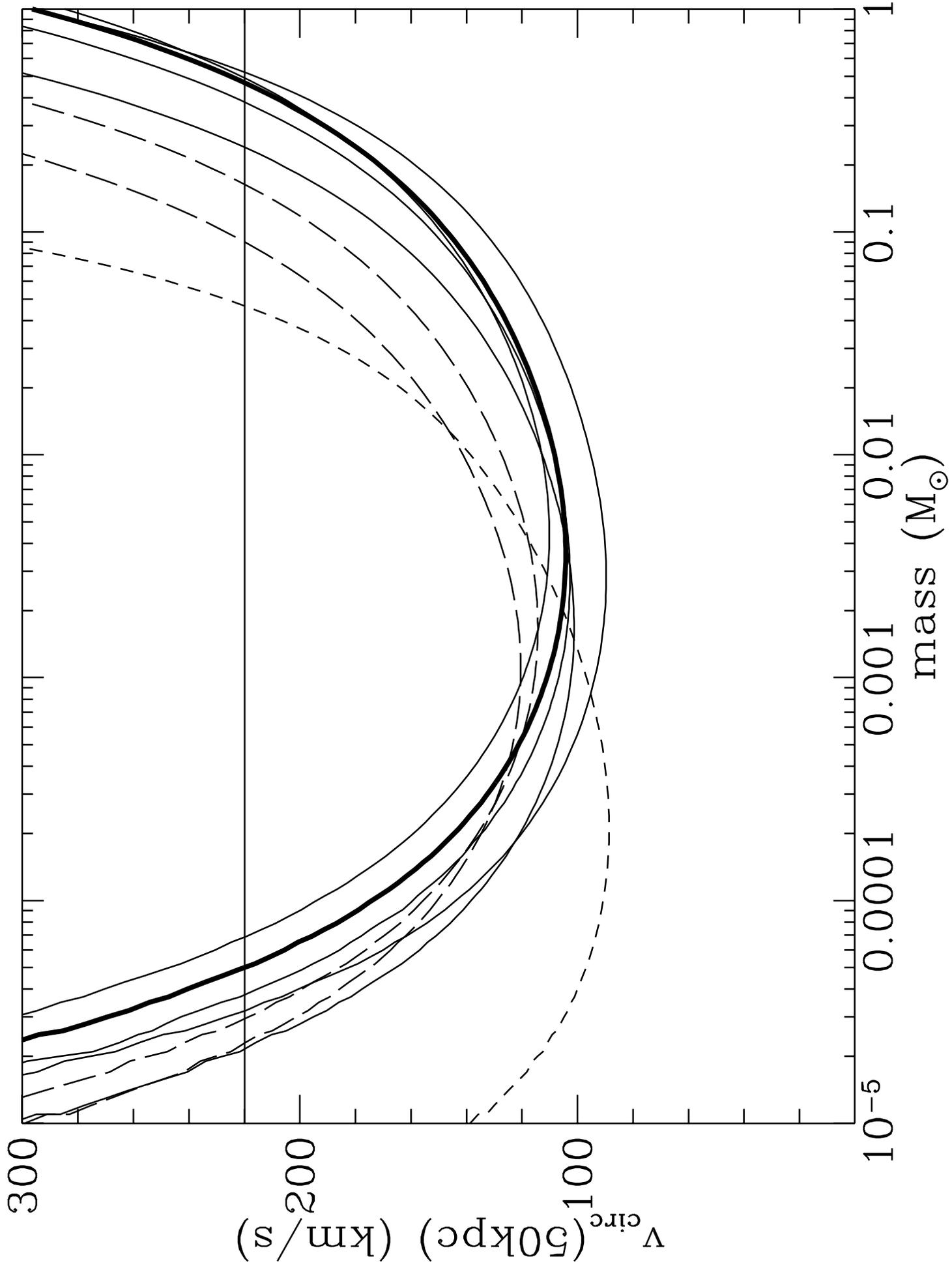

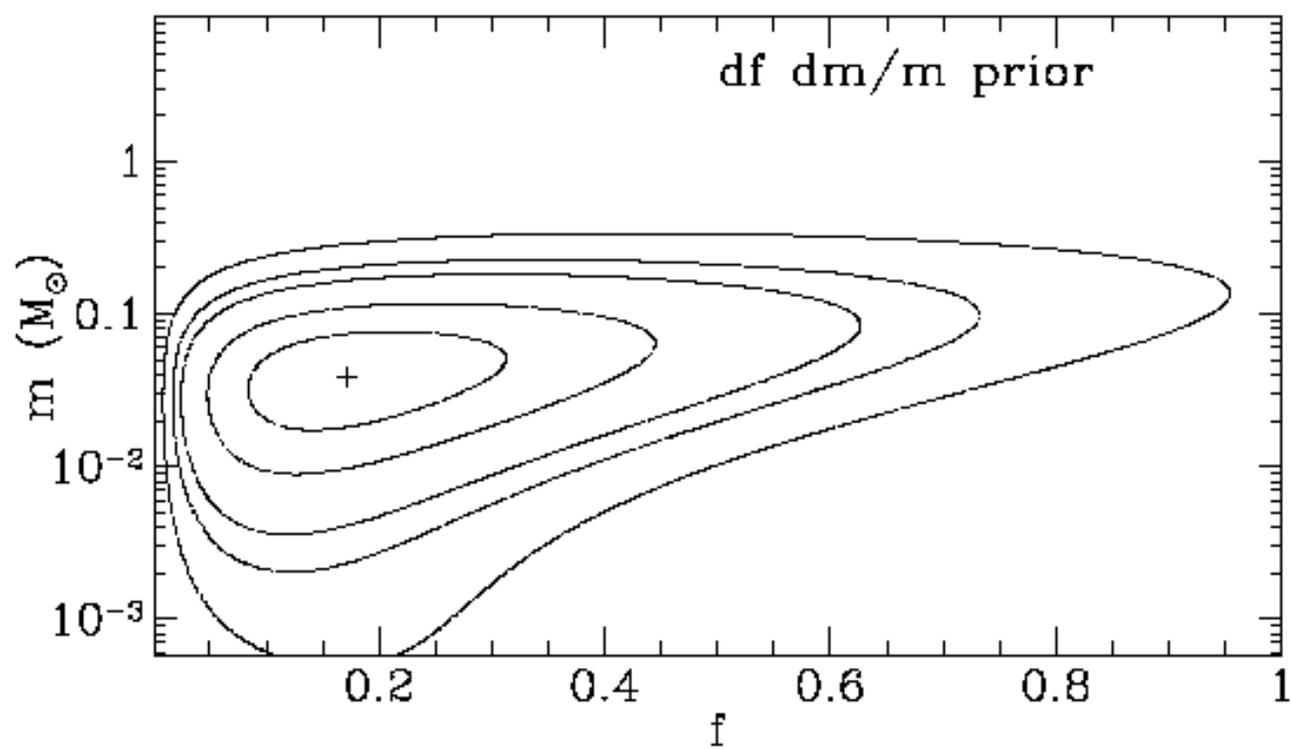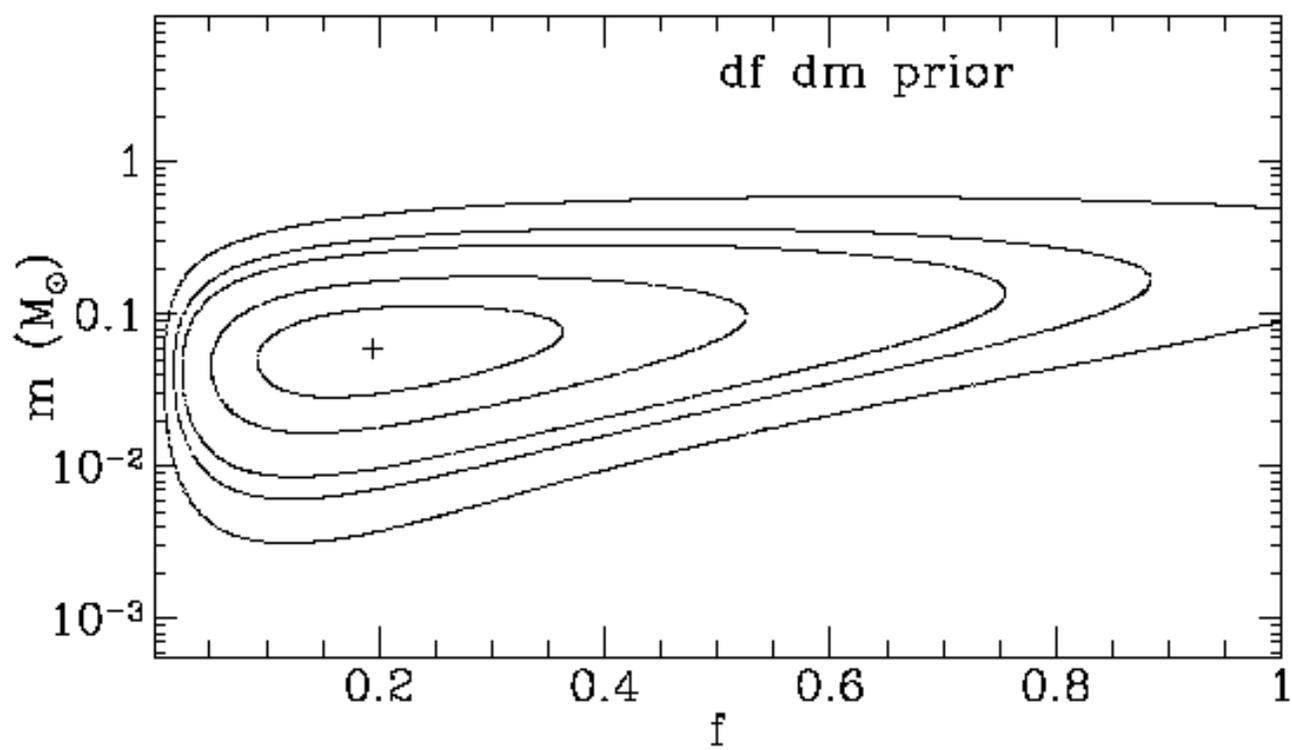

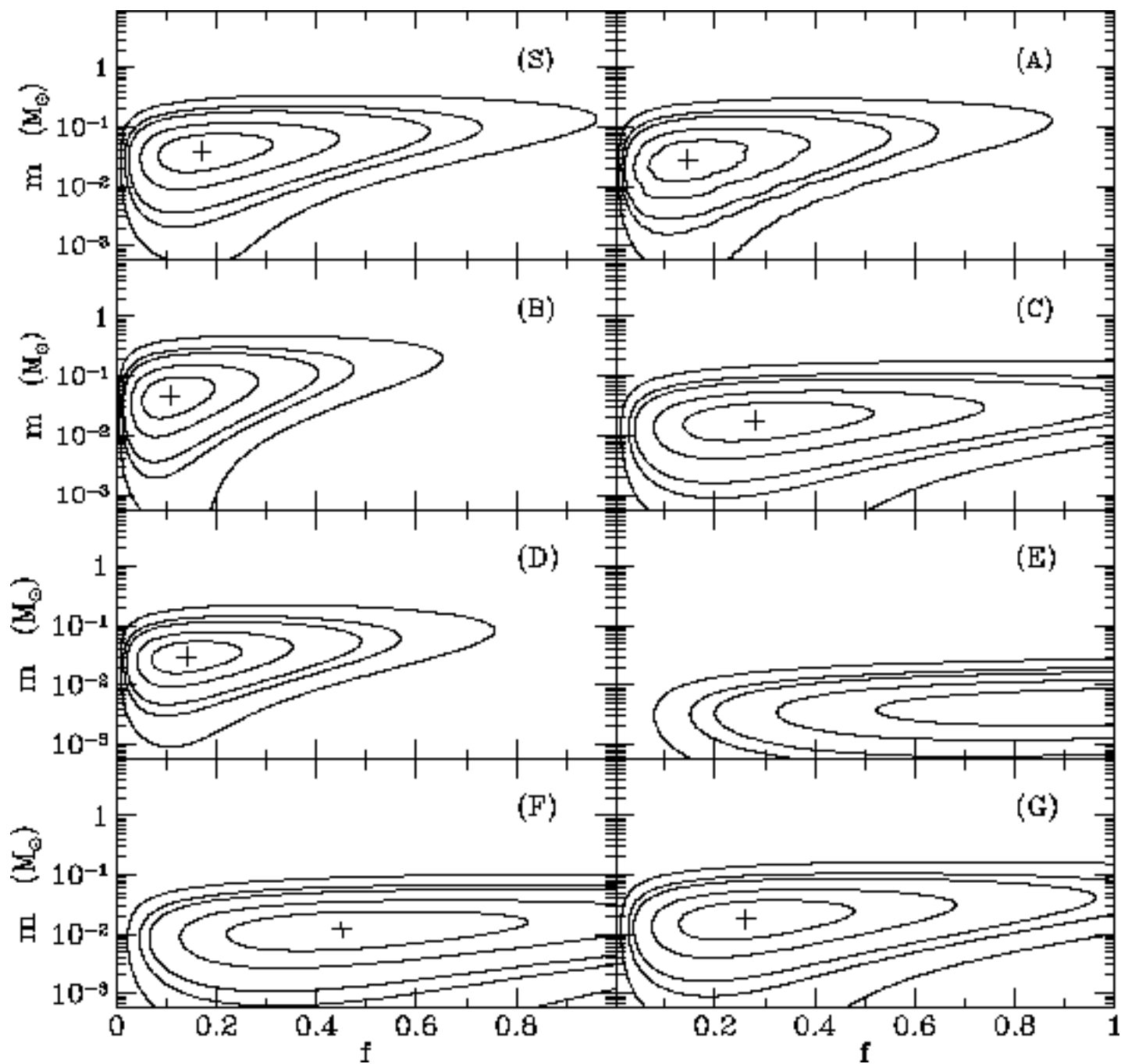